\documentclass[twocolumn,showpacs,superscriptaddress,amsmath,amssymb]{revtex4}
\usepackage{graphicx}
\usepackage{dcolumn}
\usepackage{bm}
\usepackage{color}
\usepackage{mathrsfs}
\usepackage{ulem}
\usepackage[german]{babel}

\def\bd{
\begin{document}} \def\ed{\end{document}}
\def\bmp{\begin{minipage}} \def\emp{\end{minipage}}
\def\bcc{\begin{center}} \def\ecc{\end{center}}     \def\npg{\newpage}
\def\beq{\begin{equation}} \def\eeq{\end{equation}} \def\hph{\hphantom}
\def\be{\begin{equation}} \def\ee{\end{equation}} \def\r#1{$^{[#1]}$}
\def\n{\noindent} \def\ni{\noindent} \def\pa{\parindent}
\def\hs{\hskip} \def\vs{\vskip} \def\hf{\hfill} \def\ej{\vfill\eject}
\def\cl{\centerline} \def\ob{\obeylines}  \def\ls{\leftskip}
\def\underbar#1{$\setbox0=\hbox{#1} \dp0=1.5pt \mathsurround=0pt
   \underline{\box0}$}   \def\ub{\underbar}    \def\ul{\underline}
\def\f{\left} \def\g{\right} \def\e{{\rm e}} \def\o{\over} \def\d{{\rm d}}
\def\vf{\varphi} \def\pl{\partial} \def\cov{{\rm cov}} \def\ch{{\rm ch}}
\def\la{\langle} \def\ra{\rangle} \def\EE{e$^+$e$^-$} \def\pt{p_{\rm t}}
\def\pti{p_{{\rm t},i}} \def\vti{v_{{\rm t},i}}
\def\ptj{p_{{\rm t},j}}\def\Pt{P_{\rm t}} \def\vt{v_{\rm t}}

\def\bitz{\begin{itemize}} \def\eitz{\end{itemize}}
\def\btbl{\begin{tabular}} \def\etbl{\end{tabular}}
\def\btbb{\begin{tabbing}} \def\etbb{\end{tabbing}}
\def\beqar{\begin{eqnarray}} \def\eeqar{\end{eqnarray}}
\def\\{\hfill\break} \def\dit{\item{-}} \def\i{\item}
\def\bbb{\begin{thebibliography}{9} \itemsep=-1mm}
\def\ebb{\end{thebibliography}} \def\bb{\bibitem}
\def\bpic{\begin{picture}(260,240)} \def\epic{\end{picture}}
\def\akgt{\cl{\bf ACKNOWLEDGMENTS}}
\def\fgn{\noindent{\bf\large\bf figure captions}}
\def\m1{\langle N_p\rangle} \def\u2{\langle N_{\bar p}\rangle} \def\Nap{N_{\bar
p}}
\def\lan{\langle}
\def\ran{\rangle}
\def\p{\pi}
\def\ifmath#1{\relax\ifmmode #1\else $#1$\fi}%
\def\rc{\ifmath{{\mathrm{c}}}}
\def\cut{\ifmath{{\mathrm{cut}}}}
\def\rF{\ifmath{{\mathrm{F}}}}
\def\rK{\ifmath{{\mathrm{K}}}}
\def\rp{\ifmath{{\mathrm{p}}}}
\def\rt{\ifmath{{\mathrm{t}}}}
\def\LAB{\ifmath{{\mathrm{LAB}}}}
\def\cut{\ifmath{{\mathrm{cut}}}}
\def\beq{\begin{equation}}
\def\eeq{\end{equation}}

\newcommand{\cinst}[2]{$^{\mathrm{#1}}$~#2\par}
\newcommand{\crefi}[1]{$^{\mathrm{#1}}$}
\newcommand{\crefii}[2]{$^{\mathrm{#1,#2}}$}
\newcommand{\crefiii}[3]{$^{\mathrm{#1,#2,#3}}$}
\newcommand{\HRule}{\rule{0.5\linewidth}{0.5mm}}

\bd
\title{Generalized susceptibilities of net-baryon number based on the 3-dimensional Ising universality class}

\author{Xue Pan}\email{panxue1624@163.com}
\affiliation{School of Electronic Information and Electrical Engineering, Chengdu University, Chengdu 610106, China}

\begin{abstract}
Assuming the equilibrium of the QCD system, we have investigated the critical behavior of sixth-, eighth- and tenth-order susceptibilities of net-baryon number, through mapping the results in the three-dimensional Ising model to that of QCD. Both the leading critical contribution as well as sub-leading critical contribution from the Ising model are discussed. When considering only the leading critical contribution, the density plots for susceptibilities of the same order demonstrate a consistent general pattern independent on values of mapping parameters. As the critical point is approached from the crossover side, a negative dip followed by a positive peak is observed in the $\mu_B$ dependence of the three different orders of susceptibilities. When sub-leading critical contribution is taken into account, modifications become apparent in the density plots of the susceptibilities. The emergence of negative dips in the $\mu_B$ dependence of the susceptibilities is not an absolute phenomenon, while the positive peak structure is a more robust feature of the critical point.
\end{abstract}

\pacs{25.75.Gz, 25.75.Nq}

\maketitle

\section{Introduction}

Quantum Chromodynamics (QCD) predicts that the interaction between quarks, which is strong at large separations, weakens as the quarks get closer to one another~\cite{freedom1,freedom2}. At sufficiently high temperatures or densities, a new deconfined phase of matter, quark gluon palasma (QGP), is hypothesized and expected~\cite{QGP1,QGP2}.

Lattice QCD calculations predict a crossover from hadrons to QGP at zero net-baryon chemical potential ($\mu_B$)~\cite{crossover1,crossover2}. However, due to the sign problem, the phase diagram at finite $\mu_B$ from first principles remains unknown. Some QCD based models suggest a first-order phase transition at high $\mu_B$~\cite{first1, first2, first3}. If it is true, as the decrease of $\mu_B$, the first-order phase transition line should terminate at a second-order critical point~\cite{first1, cp}.

One of the main goals of current relativistic heavy-ion collision experiments is to reveal the phase diagram of QCD, where the location of the critical point is the most important~\cite{maingoal}. The correlation length goes to infinity and the susceptibilities diverges at the critical point, which results in the non-monotonic behavior of fluctuation measures. The related observables, high-order cumulants of net-baryon number, which scale with the higher powers of the correlation length~\cite{Stephanov-prl102, Stephanov-prl107}, are a focal point for both experimental and theoretical investigations.

In the experiment, due to the fact that neutrons are uncharged, it is difficult to measure the fluctuations of net-baryon number. The alternative observables, cumulants of net-proton number, which are considered to have similar critical behavior, have been calculated up to the sixth-order~\cite{STAR1, STAR2}.

In the theory, one possible way is to extend the lattice results to finite $\mu_B$ by Taylor expansions~\cite{Taylor1,Taylor2,Taylor3} or analytical continuation from imaginary chemical potentials~\cite{imaginary1,imaginary2}. The high-order cumulants of net-baryon number have been calculated and extrapolated to small values of $\mu_B$ in Refs.~\cite{LQCD1,LQCD2,LQCD3}, but with some numerical uncertainties. An alternative method for investigating the critical behavior of high-order cumulants is QCD effective models or theories~\cite{Vladi, Chin.Phys.C.43.033103, Eur.Phys.J.C.79.245, Fuweijie-PRD104}. The other approach is based on the universality of critical behavior in phase transitions ~\cite{Stephanov-prl107,stephanov-prc103}.

If the QCD critical point exists, it should be in the same universality class with the one of the three-dimensional Ising model~\cite{class1,class2,class3,class4}. The Ising variables, reduced temperature ($t$) and magnetic field ($h$), can be mapped onto the QCD temperature and net-baryon chemical potential ($T-\mu_B$) phase plane to investigate the critical features of QCD~\cite{Asakawa-PRC71,Stephanov-PRD100,stephanov-prc101}. The $t$ axis is tangential to the QCD first-order phase transition line at the critical point. Generally, the $h$ axis will deform when mapped onto the QCD $T-\mu_B$ plane. But it is not clear how this occurs. The common assumption in existing literature is that the $h$ axis is orthogonal to the $t$ axis~\cite{Asakawa-PRC71, PRC92.034912}.

A sketch of mapping the Ising variables onto the QCD $T-\mu_B$ phase plane is shown in Fig. 1. The solid black line represents the QCD first-order phase transition line. The red point is the QCD critical point. $\alpha_1$ and $\alpha_2$ represent the angles between the horizontal axis (where $T$ is a constant) and $t$ axis and $h$ axis in the Ising model when they are mapped onto the QCD $T-\mu_B$ phase plane, respectively.

\begin{figure}[hbt]
	\centering
	\includegraphics[width=0.35\textwidth]{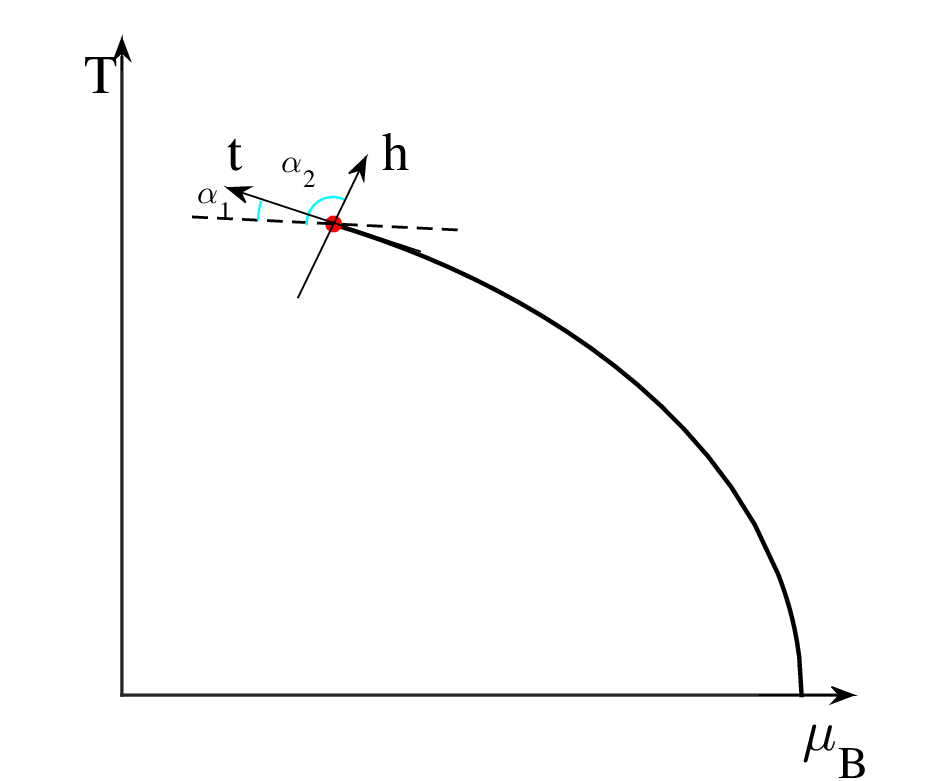}
	\caption{\label{Fig. 1}(Color online). A sketch of mapping the Ising temperature $t$ and magnetic field $h$ onto the QCD $T-\mu_B$ phase plane. The red point represents the QCD critical point at the end of the first-order phase transition line.}
\end{figure}

When considering the critical behavior of cumulants of net-baryon number, usually only the leading critical contribution from the Ising model has been taken into account~\cite{Stephanov-prl107,NPA}. When the critical point is approached along the $h$ axis and $t$ axis, the baryon-baryon correlation length diverges with the exponent $y_h$ and $y_t$, respectively~\cite{Gupta-PLB696}. $y_h\approx2.5$ is bigger than $y_t\approx1.6$~\cite{Physica A326}. So the cumulants of magnetization (order parameter in the Ising model) should dominate the critical behavior of cumulants of net-baryon number.

In particular, on this basis, it is predicted that as the critical point is approached from the crossover side, the fourth-order cumulant of net-baryon number has a negative dip~\cite{Stephanov-prl107}. The negative dip has been regarded as a critical signal and used to locate the QCD critical point in experiments~\cite{STAR3}.

However, it is pointed out that the sub-leading critical contribution can significantly affect the $\mu_B$ dependence of the cumulants of net-baryon number along the freeze-out curve, even the negative dip in the fourth-order cumulant disappears in the case that $\alpha_2-\alpha_1=90^\circ$~\cite{stephanov-prc103}.

In our earlier literature, the negative dip in the sixth-order cumulant of the magnetization in the three-dimensional Ising model is predicted when the critical point is approached from the crossover side~\cite{cpc2016}. It is interesting to map the results of the high-order cumulants in the Ising model to QCD, and study the influence of the sub-leading singular contribution.

On the other hand, recent lattice results predict that the QCD critical temperature is lower than $135$ MeV~\cite{chiraltransitionT1,chiraltransitionT2}. What is more, it has been shown within the functional renormalization group and Dyson-Schwinger equations approaches that, the transition from hadronic phase to QGP is a crossover with increasing $\mu_B$ for $\mu_B/T \lesssim 4$~\cite{Fuweijie-PRD101,Fischer,Isserstedt,Gaofei1,Gaofei2}. And a QCD critical point is found at larger $\mu_B$, although beyond the quantitative reliability of the theory computations~\cite{Fuweijie-PRD101,Gaofei1,Gaofei2}. So studying the behavior of high-order cumulants of net-baryon number at large $\mu_B$ for $\mu_B/T > 4$ is necessary.

In this paper, we assume the equilibrium of the QCD system and the existence of QCD critical point at $(T_C,\mu_{BC})=(107, 635)$ MeV~\cite{Fuweijie-PRD101}, where $T_C$ and $\mu_{BC}$ are the temperature and net-baryon chemical potential at the QCD critical point, respectively. Through mapping the results from the three-dimensional Ising model to that of QCD, we study the critical behavior of sixth-, eighth- and tenth-order susceptibilities of net-baryon number. They have similar behavior with the corresponding cumulants. The leading as well as sub-leading critical contribution from the Ising model is discussed.

The paper is organized as follows. In section 2, the parametric representation of the three-dimensional Ising model is introduced. Furthermore, the linear mapping from this model to QCD is presented and the expression of generalized susceptibilities of net-baryon number is deduced. In section 3, the effects of leading singular contribution on the behavior of the sixth-, eighth- and tenth-order susceptibilities of net-baryon number is analyzed and discussed, considering different values of mapping parameters. In section 4, effects of the sub-leading singular contribution on the behavior of these susceptibilities is investigated and discussed. Finally, conclusions and summary are given in section 5.

\section{The linear mapping from Ising model to QCD}
In the parametric representation of the three-dimensional Ising model, magnetization ($M$) and reduced temperature ($t$) can be parameterized by two variables $R$ and $\theta$~\cite{linearpara, linearpara3},
\begin{equation}\label{parametric}
	M=m_0R^{\beta}\theta,~~~~~~t=R(1-\theta^2).
\end{equation}
The equation of state expressed by $R$ and $\theta$ is
\begin{equation}\label{equation state}
	h=h_0R^{\beta\delta}\widetilde h(\theta).
\end{equation}
Where $m_0\simeq0.605$ in Eq.~\eqref{parametric} and $h_0\simeq0.364$ in Eq.~\eqref{equation state} are normalization constants. They are fixed by imposing the normalization conditions $M(t=-1,h=+0)=1$ and $M(t=0,h=1)=1$. $\beta\simeq0.326$ and $\delta\simeq4.8$ are critical exponents of the three-dimensional Ising universality class~\cite{Isingexponents}. $\widetilde h(\theta)=\theta(1-0.76201\theta^{2}+0.00804\theta^{4})$. The parameters are within the range $R\geqslant0$ and $|\theta|\leqslant1.154$.

The free energy density can be defined as~\cite{stephanov-prc101},
\begin{equation}\label{free energy density}
	F(M,t)=h_0m_0R^{2-\alpha}g(\theta),
\end{equation}
where $\alpha\simeq0.11$ is another critical exponent of the three-dimensional Ising universality class. The relation $2-\alpha=\beta(\delta-1)$ holds, and
\begin{equation}\label{gtheta}
	g(\theta)=c_0+c_1(1-\theta^2)+c_2(1-\theta^2)^2+c_3(1-\theta^2)^3,
\end{equation}
with
\begin{equation}
	\begin{split}
		&c_0=\frac{\beta}{2-\alpha}(1+a+b),\\
		&c_1=-\frac{1}{2(\alpha-1)}[(1-2\beta)(1+a+b)-2\beta(a+2b)],\\
		&c_2=-\frac{1}{2\alpha}[2\beta b-(1-2\beta)(a+2b)],\\
		&c_3=-\frac{1}{2(\alpha+1)}b(1-2\beta).
		\nonumber
	\end{split}
\end{equation}
Then the Gibbs free energy density is
\begin{equation}\label{Gibbs free energy density}
	G(h,t)=F(M,t)-Mh.
\end{equation}

The pressure equals to the Gibbs free energy density up to a minus sign: $P = -G$, and hence the pressure in the Ising model can be written as follows
\begin{equation}\label{Ising pressure}
	P^{Ising}(R,\theta)=h_0m_0R^{2-\alpha}[\theta\widetilde h(\theta)-g(\theta)].
\end{equation}

The $n_{th}$-order susceptibility of the magnetization and energy represented by $R$ and $\theta$ can be got from the derivatives of the pressure with respect to $h$ and $t$,
\begin{equation}\label{cumulant of Ising}
	\chi_{n_1,n_2}^{M,E}=\frac{\partial^{n_1+n_2}P^{Ising}}{\partial h^{n_1}\partial t^{n_2}}, n=n_1+n_2.
\end{equation}
When $n_2=0$, it is the $n_{th}$-order susceptibility ($\chi_{n}^{M}$) of the magnetization. When $n_1=0$, it is the $n_{th}$-order susceptibility ($\chi_{n}^{E}$) of the energy.

For examples,
\begin{equation}\label{cumulant of M}
	\begin{split}
		\chi_1^{M} &= \left(\frac{\partial P^{Ising}}{\partial h} \right)_t \\
		&=\frac{\partial P^{Ising}}{\partial R}\left(\frac{\partial R}{\partial h} \right)_t+\frac{\partial P^{Ising}}{\partial \theta}\left(\frac{\partial \theta}{\partial h}\right)_t,
	\end{split}
\end{equation}
\begin{equation}\label{cumulant of E}
	\begin{split}
		\chi_1^{E} &= \left(\frac{\partial P^{Ising}}{\partial t} \right)_h  \\
		&=\frac{\partial P^{Ising}}{\partial R}\left(\frac{\partial R}{\partial t} \right)_h+\frac{\partial P^{Ising}}{\partial \theta}\left(\frac{\partial \theta}{\partial t}\right)_h.
	\end{split}
\end{equation}
Where $\partial R /\partial h$, $\partial \theta /\partial h$, $\partial R /\partial t$, and $\partial \theta /\partial t$ can be got from Eqs.~\eqref{parametric} and \eqref{equation state}.

In order to map the results of the Ising model to that of QCD, a linear relationship~\cite{linearmap1, stephanov-prc101, stephanov-prc103} including six mapping parameters can be written as follows:
\begin{equation}\label{linear map}
	\frac{T-T_C}{T_C}=w(\rho t\sin\alpha_1+h\sin\alpha_2),
\end{equation}
\begin{equation}\label{linear map1}
	\frac{\mu_B-\mu_{BC}}{T_C}=w(-\rho t\cos\alpha_1-h\cos\alpha_2),
\end{equation}
where $w$ and $\rho$ are two scaling parameters of the mapping from Ising model to QCD. $\alpha_1$ and $\alpha_2$ are two angles which have been introduced in Section 1.

In order to reduce the number of mapping parameters, supposing the QCD phase transition line up to $O(\mu_B^4)$ is expressed as follows,
\begin{equation}\label{QCD transition line}
	T= T_0[1-\kappa(\frac{\mu_B}{T_0})^2-\lambda (\frac{\mu_B}{T_0})^4],
\end{equation}
where $T_0$ is the transition temperature at $\mu_B=0$, which is set as $156.5$ MeV based on the lattice results in Ref.~\cite{Bazavov2019}. $\kappa$ is the curvature at $\mu_B=0$, which is set as $0.015$. Recent results from the functional renormalization group and Dyson-Schwinger equations approaches~\cite{Fuweijie-PRD101,Gaofei1,Gaofei2} and lattice QCD~\cite{fodor-plb751,Bazavov2019,Borsanyi2020} agree within the errors with $\kappa \approx 0.015$. Based on the results in Refs.~\cite{Fuweijie-PRD101,Fischer,Isserstedt,Gaofei1,Gaofei2}, the transition from hadronic phase to QGP is a crossover with increasing $\mu_B$ for $\mu_B/T \lesssim 4$. The QCD critical point might exist beyond this region~\cite{Fuweijie-PRD101,Fischer,Isserstedt,Gaofei1,Gaofei2}. In this paper, we adopt the critical point $(T_C,\mu_{BC})=(107, 635)$ MeV in Ref.~\cite{Fuweijie-PRD101}. In order to let the QCD phase transition line go through the critical point, $\lambda$ is set as $0.000256$. This value is in agreement with the lattice results within the errors in Refs.~\cite{Bazavov2019, Borsanyi2020}. 

Because the $t$ axis in the Ising model is tangential to the first-order phase transition line at the QCD critical point, the value of $\alpha_1$ can be fixed at $10.8^{\circ}$. For the common 'default' choice in the literature, the $h$ axis is orthogonal to the $t$ axis. So we set $\alpha_2=100.8^\circ$ in this paper. Then only two mapping parameters $w$ and $\rho$ are unknown.

\begin{figure*}[hbt]
	\centering
	\includegraphics[width=0.32\textwidth]{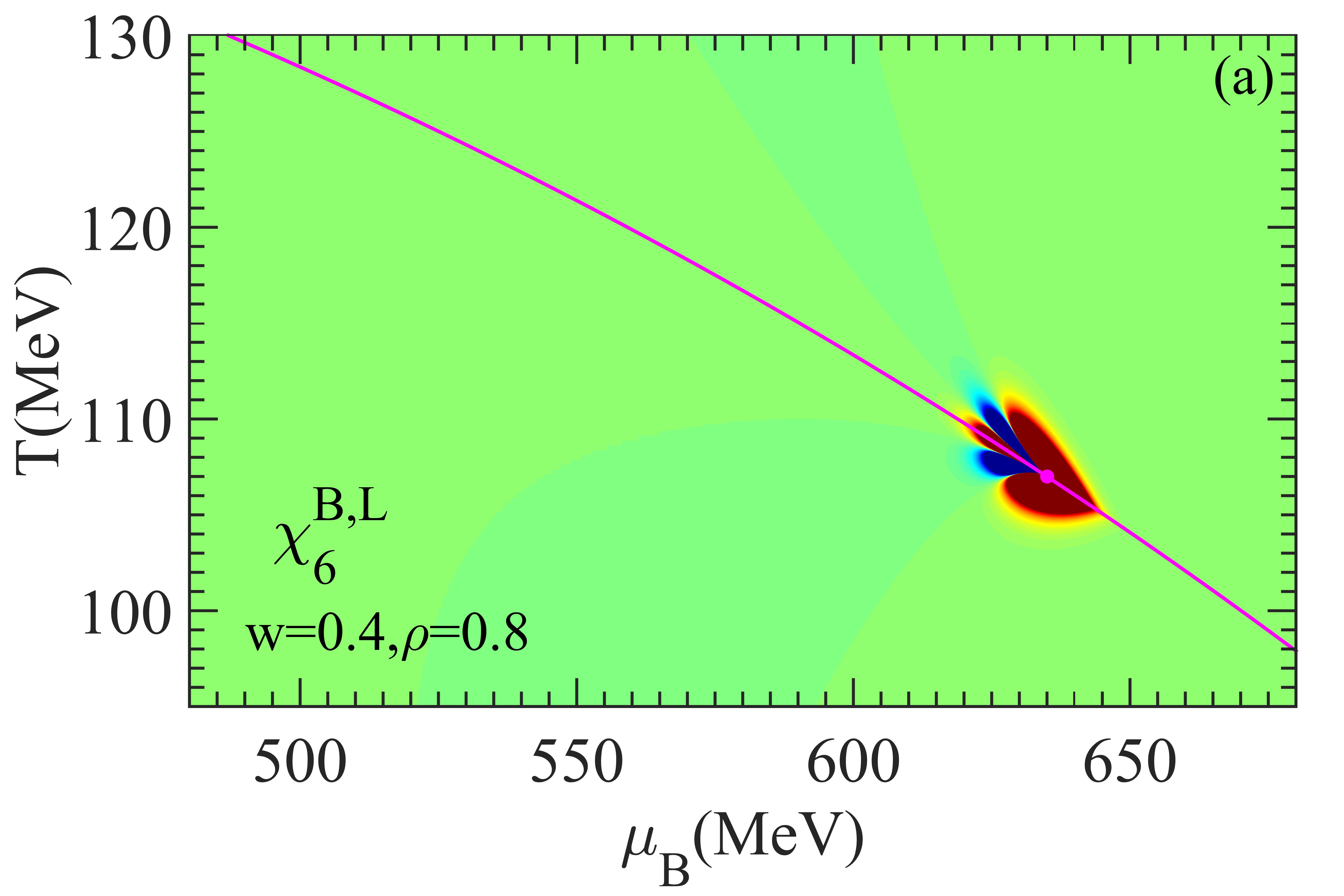}
	\includegraphics[width=0.32\textwidth]{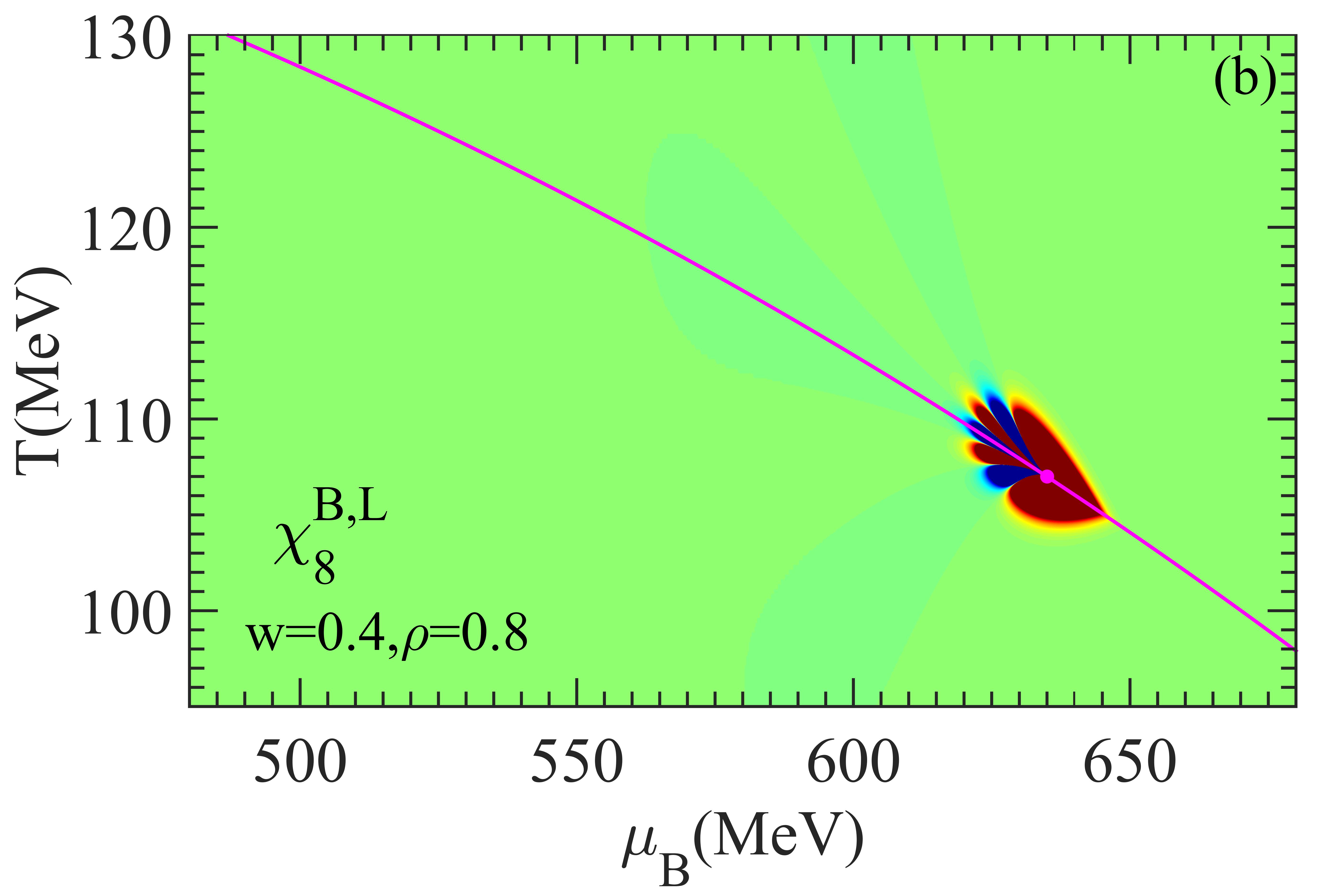}
	\includegraphics[width=0.32\textwidth]{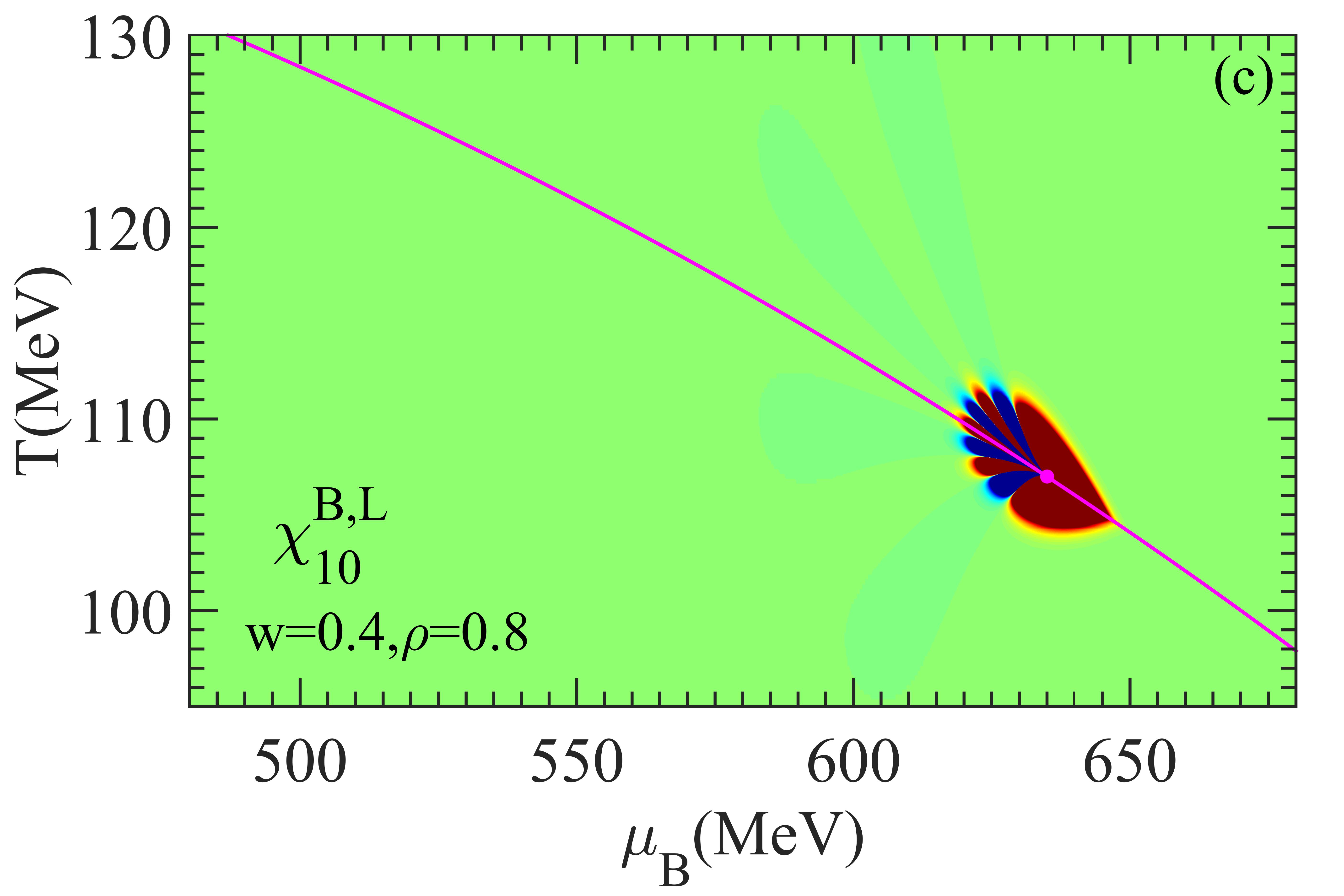}
	\includegraphics[width=0.32\textwidth]{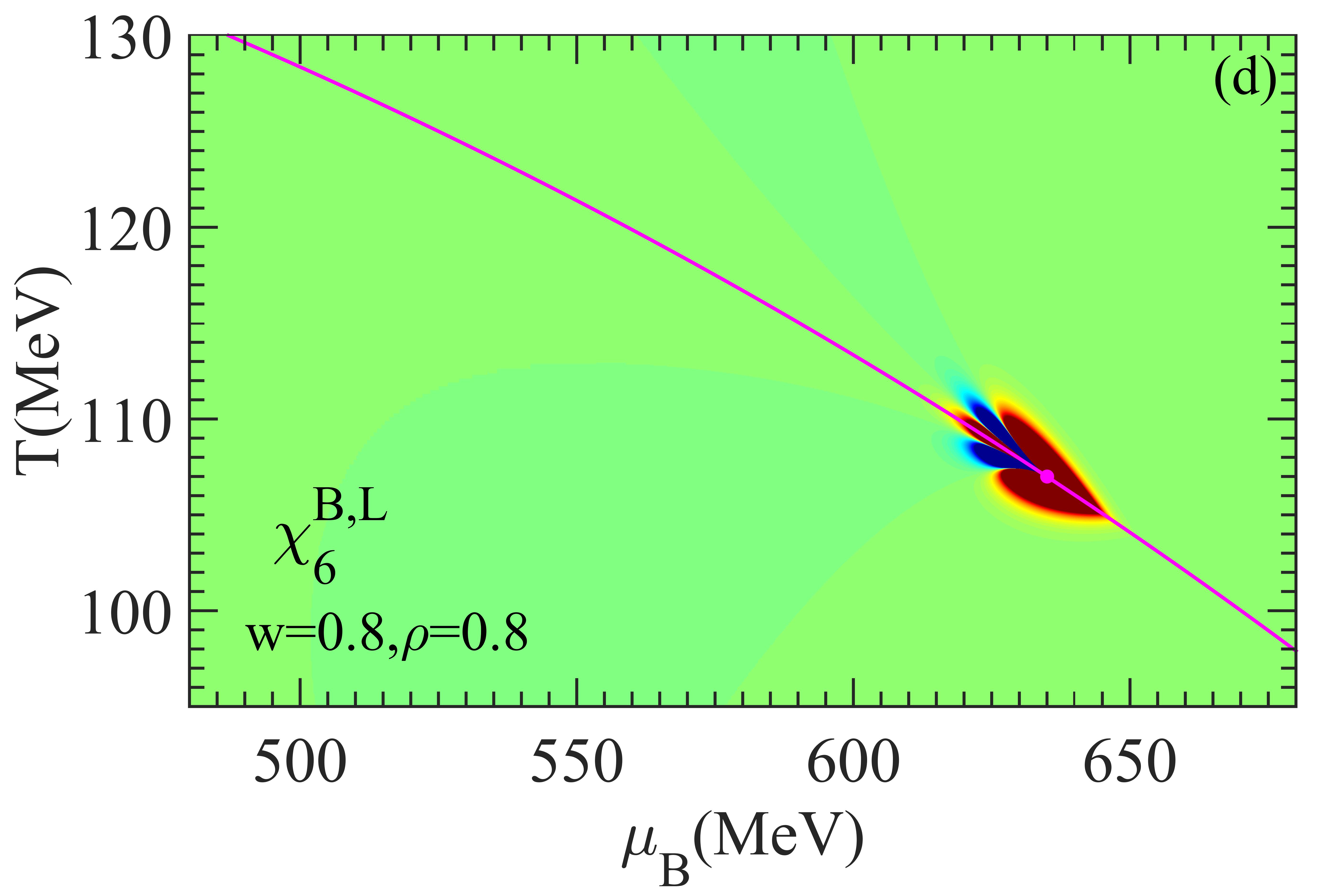}
	\includegraphics[width=0.32\textwidth]{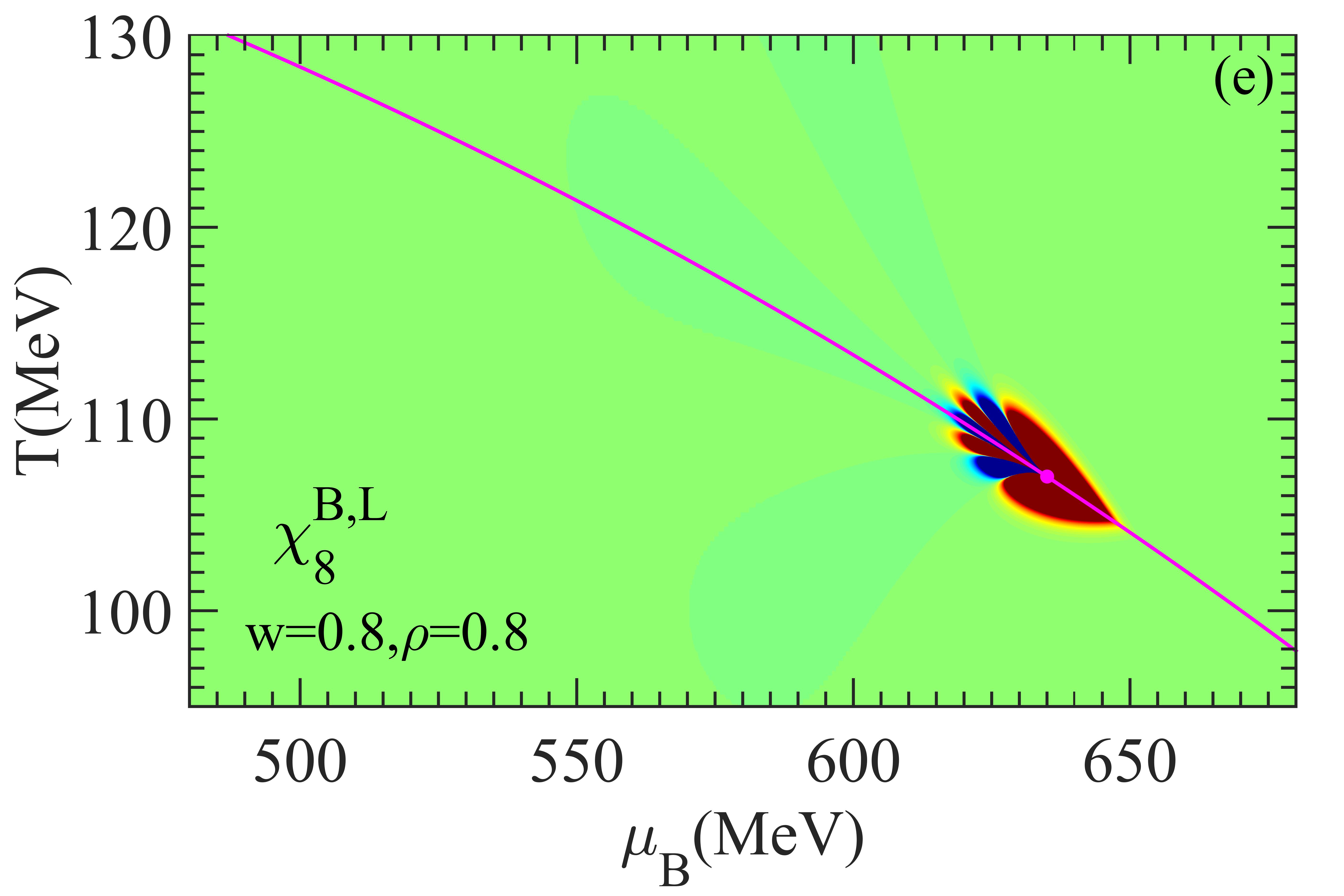}
	\includegraphics[width=0.32\textwidth]{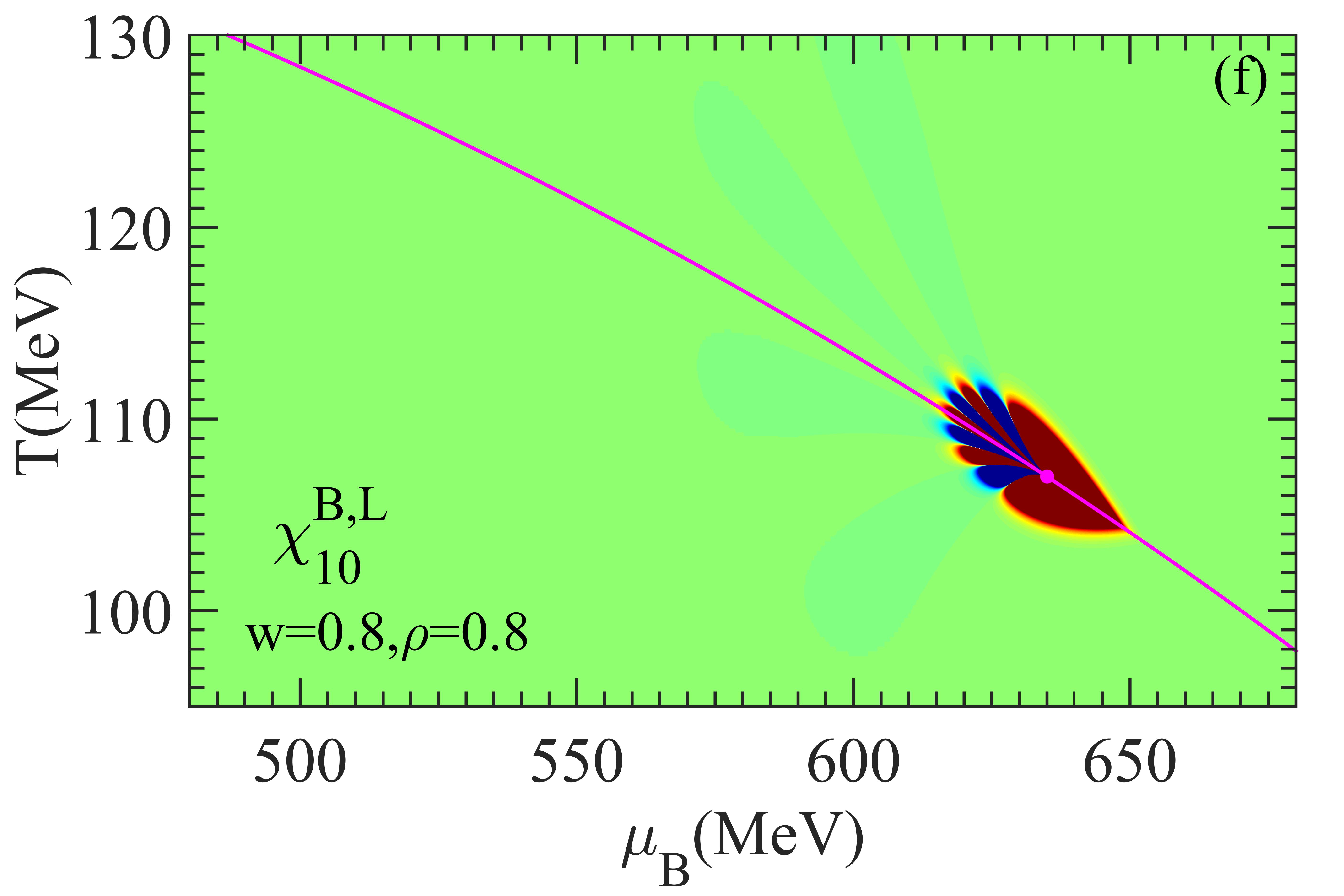}
	\includegraphics[width=0.32\textwidth]{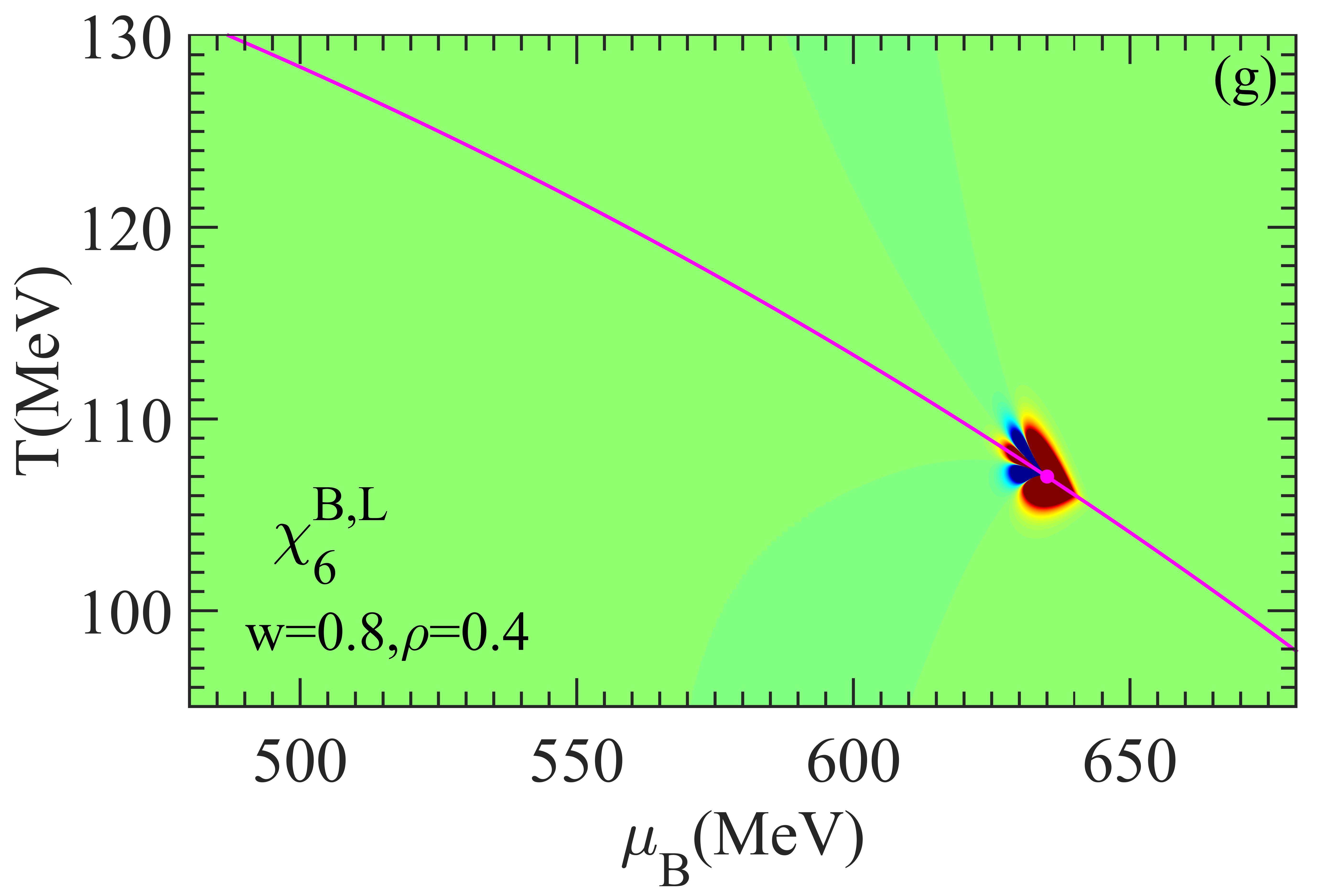}
	\includegraphics[width=0.32\textwidth]{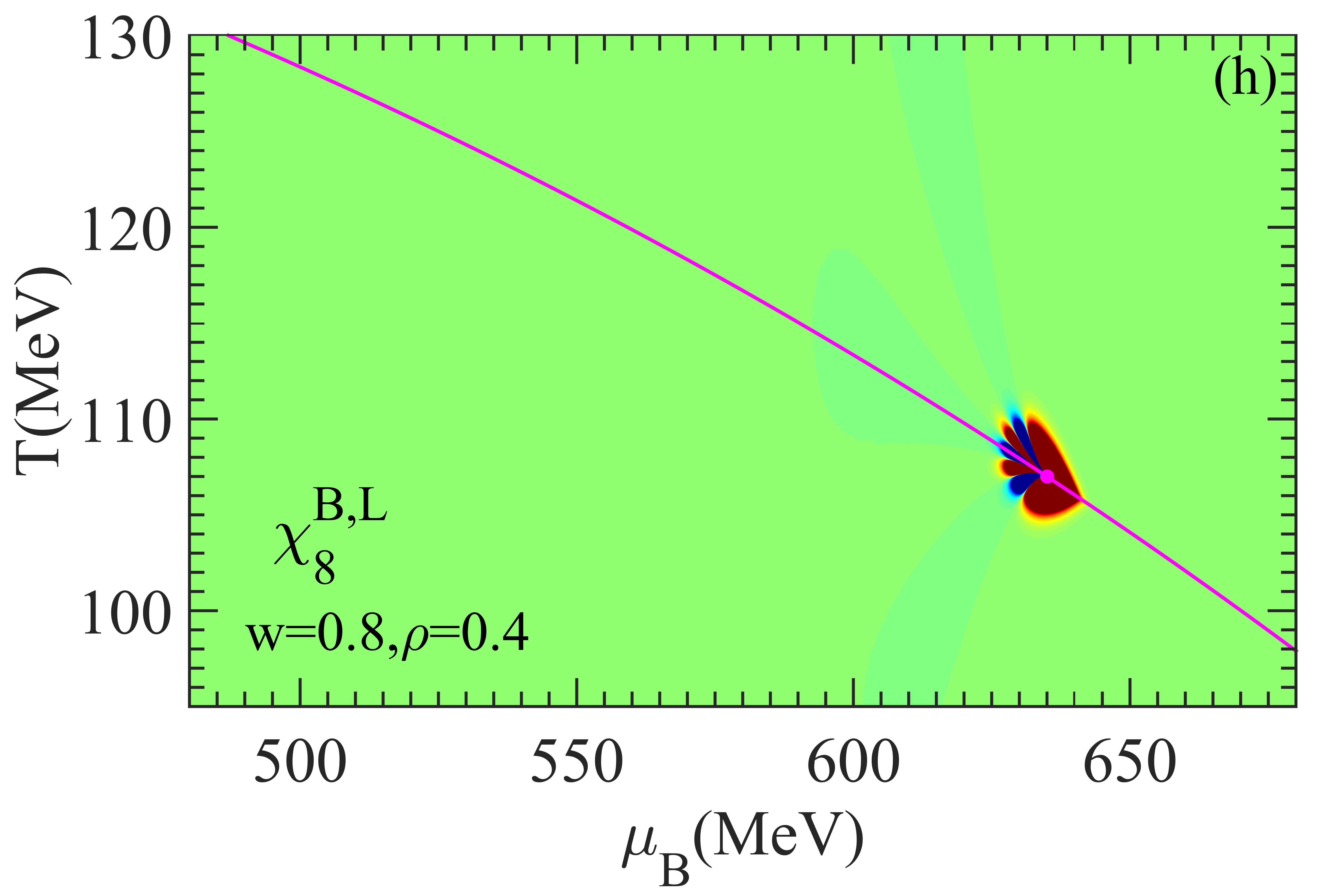}
	\includegraphics[width=0.32\textwidth]{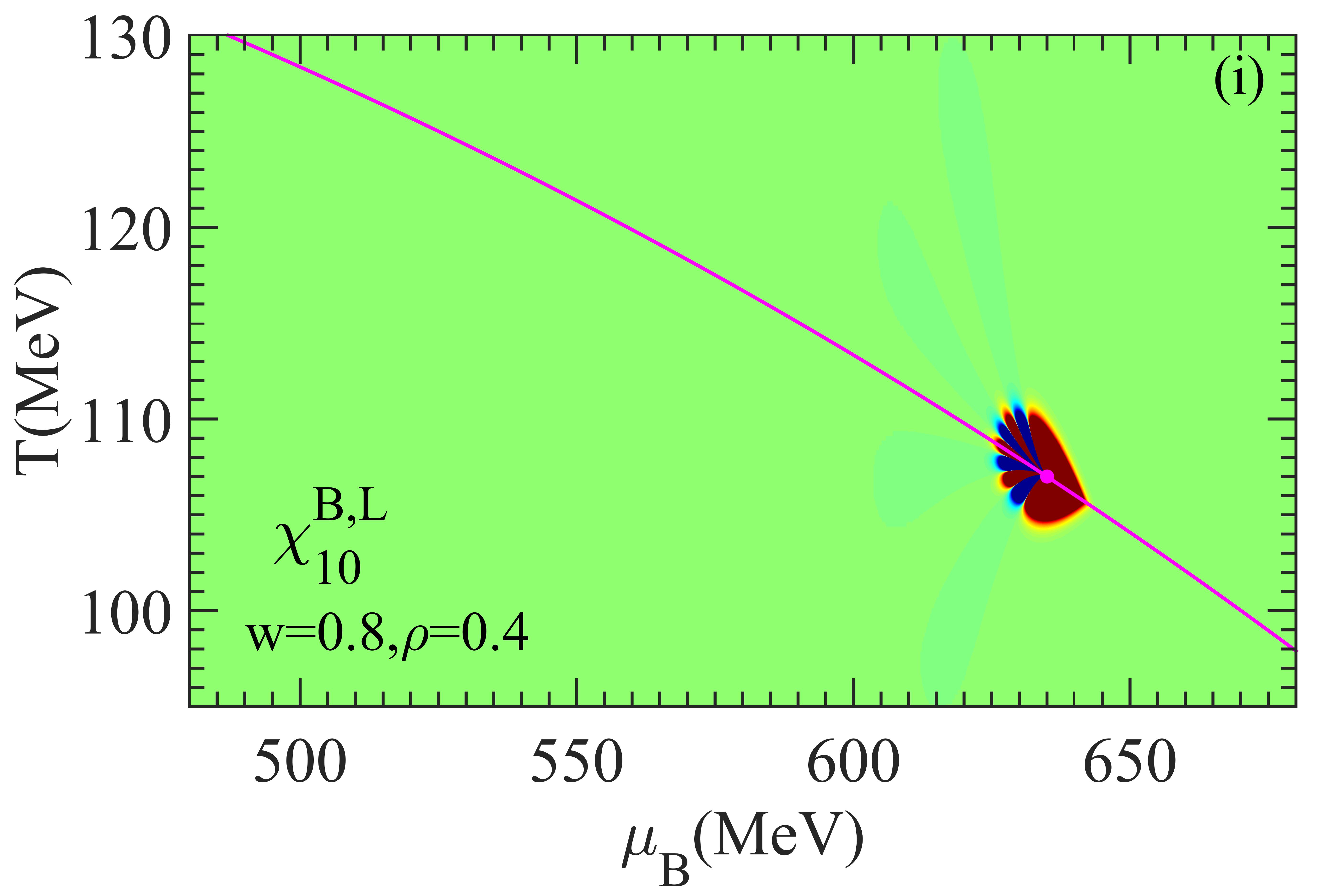}
	\caption{\label{Fig. 2}(Color online). Density plots of critical contribution to $\chi_{6}^{B,L}$, $\chi_{8}^{B,L}$, and $\chi_{10}^{B,L}$ in the QCD $T-\mu_B$ phase plane with $w=0.4, \rho=0.8$ (top row), $w=0.8, \rho=0.8$ (middle row) and $w=0.8, \rho=0.4$ (bottom row). The critical point is indicated by a purple dot, while the chiral phase transition line is represented by the solid purple line. The green, yellow and red areas correspond to positive values (the regions where it is the largest and smallest are indicated in red and green, respectively) of the susceptibilities, while the blue ones correspond to negative values (the darker the blue, the larger in magnitude of the susceptibilities).}
\end{figure*}

Generalized susceptibilities of net-baryon number ($\chi_n^{B}$) can be obtained from the $n$th-order derivatives of the pressure with respect to $\mu_B$ at fixed $T$:
\begin{equation}\label{net-baryon number susceptibility}
	\chi_n^{B}(T,\mu_B)=\left(\frac{\partial^{n} P/T^4}{\partial (\mu_B/T)^n}\right)_T.
\end{equation}

The full QCD pressure can be reconstructed as Ref.~\cite{stephanov-prc101},
\begin{flalign}\label{pressure}
	& \qquad P(T,\mu_B)=T^4 \sum_{n}c_n^{Non-Ising}(T)\left(\frac{\mu_B}{T}\right)^n &  \nonumber \\
	& \qquad \qquad  +P_C^{QCD}(T,\mu_B),&
\end{flalign}
where the first term on the left side is the Taylor expansion of the pressure from the "Non-Ising" contribution. $c_n^{Non-Ising}(T)$ is the corresponding Taylor expansion coefficients. While $P_C^{QCD}(T,\mu_B)$ represents the critical pressure mapped from the three-dimensional Ising model onto QCD. The details can be got from Refs.~\cite{stephanov-prc101, stephanov-prc103}.

In this paper, we only consider the critical point contribution to the behavior of the sixth-, eighth- and tenth-order susceptibilities of net-baryon number, the pressure in Eq.~\eqref{net-baryon number susceptibility} can be written as follows~\cite{stephanov-prc101},
\begin{equation}\label{QCD critical pressure}
	P(T,\mu_B)=T_C^4P^{Ising}(R(T,\mu_B),\theta(T,\mu_B)).
\end{equation}

The $2n$th-order susceptibility of net-baryon number can be written as
\begin{flalign}\label{susceptibilities}
	& \chi_{2n}^B=T_C^4T^{2n-4}\times &  \nonumber \\
	& \sum_{k=0}^{2n} C(2n,k) \left(\frac{\partial h}{\partial \mu_B}\right)^{2n-k} \left(\frac{\partial t}{\partial \mu_B}\right)^{k}\chi_{2n-k,k}^{M,E}, &
\end{flalign}
where $C(2n,k)=(2n)!/k!/(2n-k)!$. $\partial h /\partial \mu_B$ and $\partial t /\partial \mu_B$ can be got from Eq.~\eqref{linear map1},
\begin{flalign}\label{partial h}
	& \partial h /\partial \mu_B=-\sin(\alpha_1)/(T_c w \sin(\alpha_1 - \alpha_2)), &  \nonumber \\
	& \partial t /\partial \mu_B=\sin(\alpha_2)/(T_c w \rho \sin(\alpha_1 - \alpha_2)).
\end{flalign}

If only considering the leading singular contribution, i.e. $k$ just take the value $0$ in Eq.~\eqref{susceptibilities}, the corresponding $2n$th-order susceptibility of net-baryon number is as follows,
\begin{flalign}\label{six}
	\chi_{2n}^{B,L}=T_C^4T^{2n-4}\left(\frac{\partial h}{\partial \mu_B}\right)^{2n}\chi_{2n}^{M}.
\end{flalign}

\section{Leading critical contribution to the behavior of susceptibilities of net-baryon number}

When only considering the leading critical contribution of the mapping from the Ising model to QCD, density plots of the sixth-, eighth- and tenth-order susceptibilities of net-baryon number are shown in the left, middle, and right column of Fig.~2, respectively. The values of mapping parameters are the same for each row. That is $w=0.4$, $\rho=0.8$ for the top row, $w=0.8$, $\rho=0.8$ for the middle row, and $w=0.8$, $\rho=0.4$ for the bottom row.

The color function of the three columns of Fig.~2 is different because of the big difference of the magnitude of different orders of susceptibilities. The color schemes are such that a factor one thousands in the value of $\chi_8^{B,L}$ and $1$ million in the value of $\chi_{10}^{B,L}$ separates the middle and right columns with the left column, for the same color.

In each sub-figure, the green, yellow and red areas correspond to positive values. The regions where the value is largest and smallest are indicated in red and green, respectively. The blue areas correspond to negative values, and the darker, the larger in its magnitude. The purple curve shows the QCD phase transition line represented by Eq.\eqref{QCD transition line}. The purple dot marks the critical point.

\begin{figure*}[hbt]
	\centering
	\includegraphics[width=0.32\textwidth]{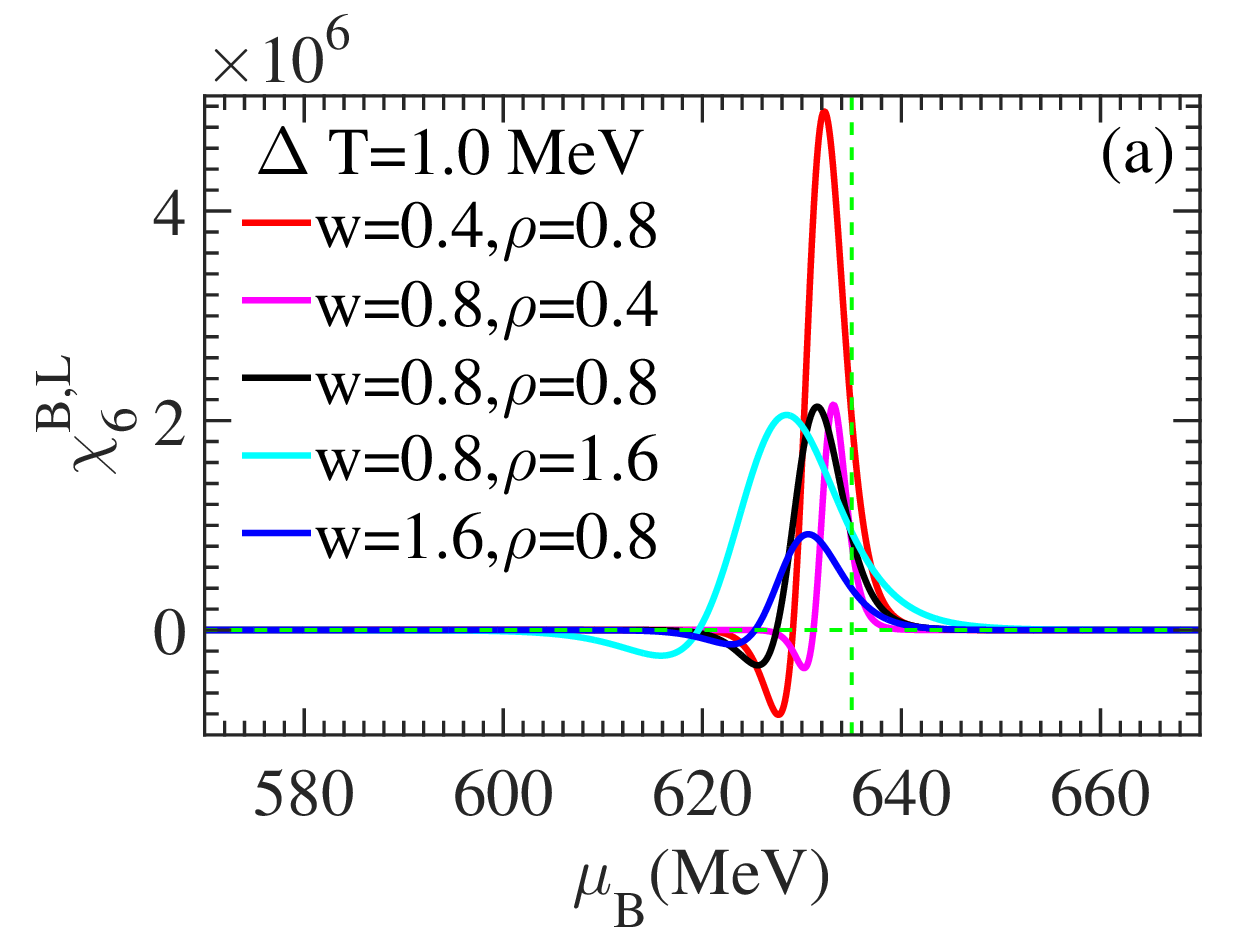}
	\includegraphics[width=0.32\textwidth]{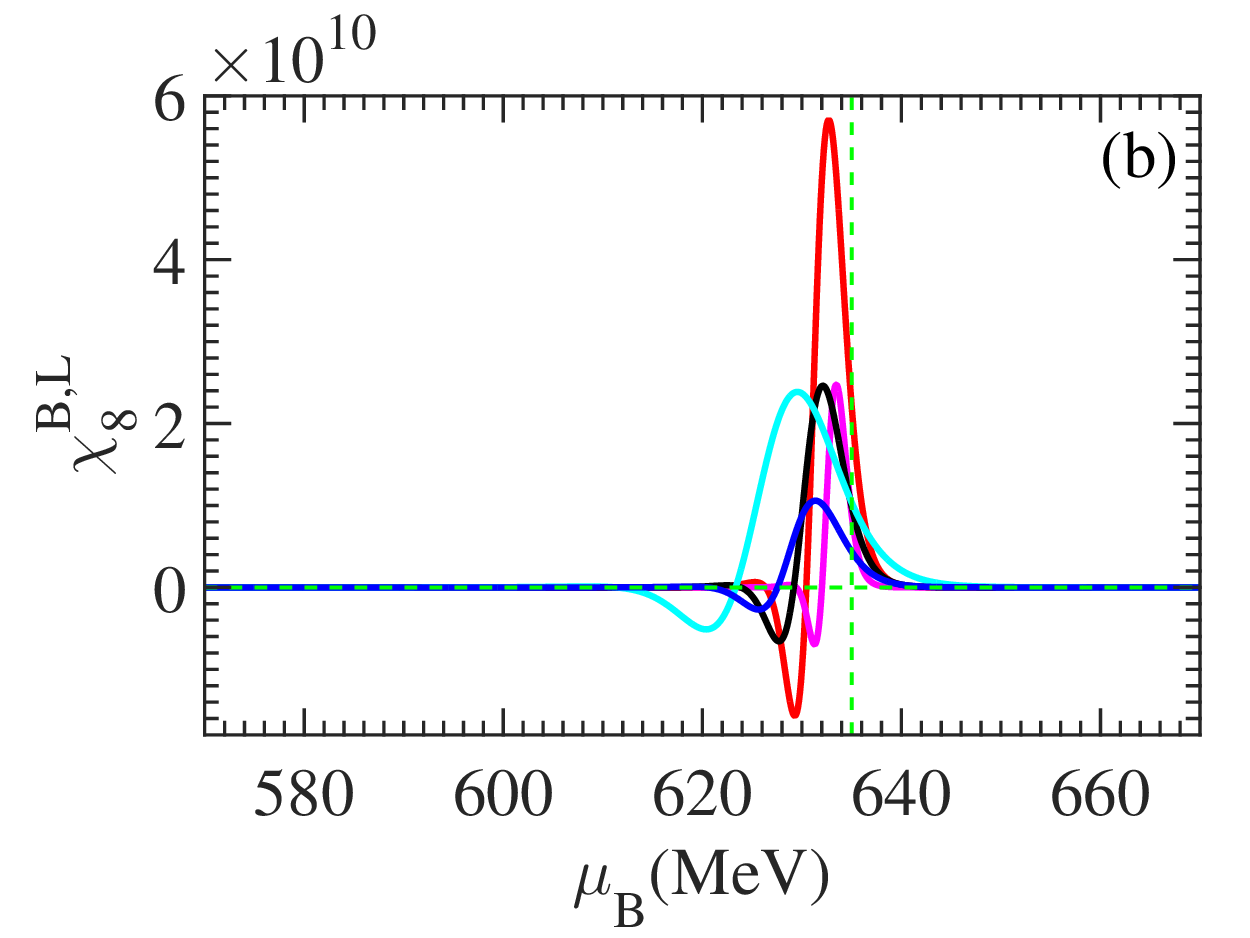}
	\includegraphics[width=0.32\textwidth]{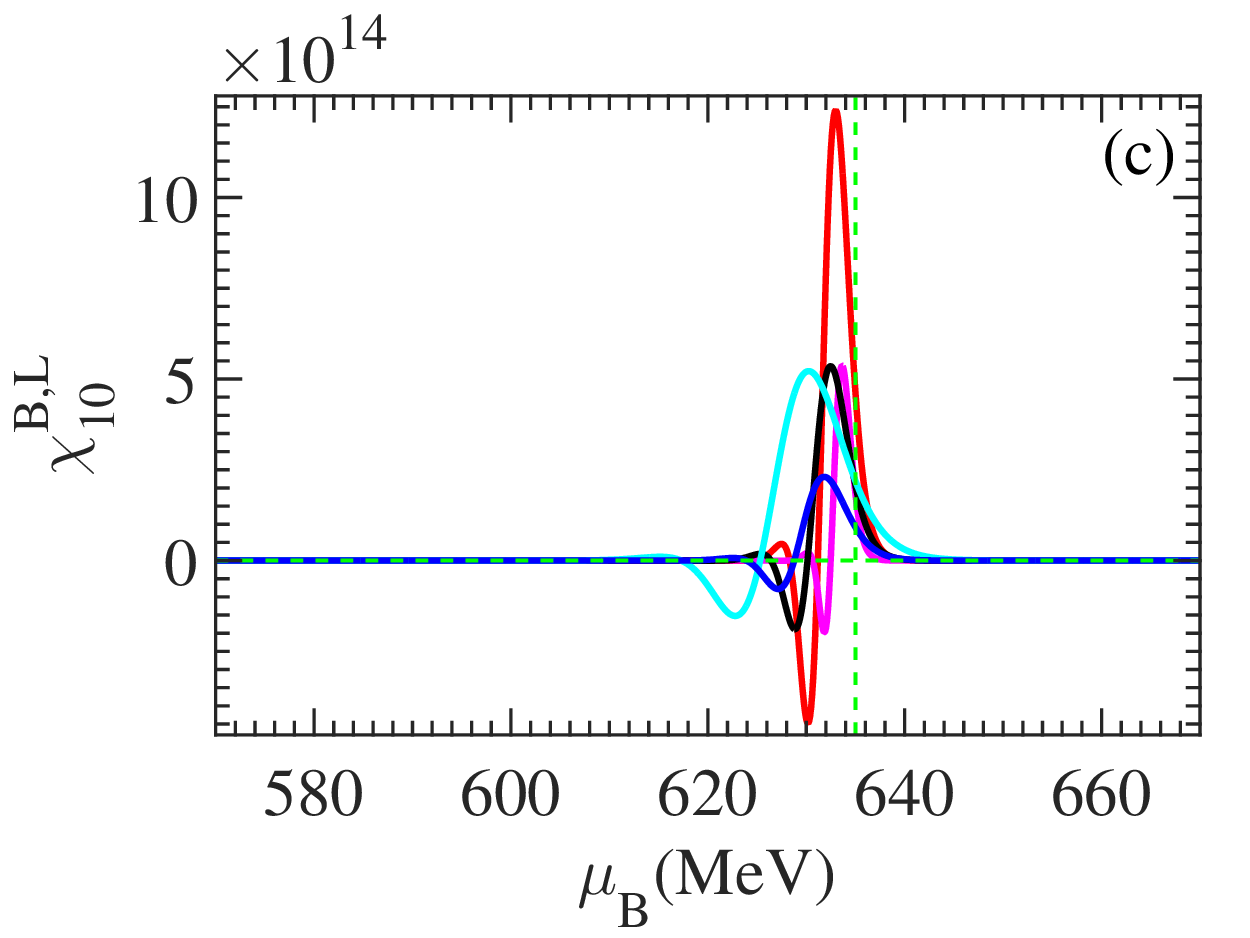}
	\caption{\label{Fig. 3}(Color online). $\mu_B$ dependence of $\chi_{6}^{B,L}$ (a), $\chi_{8}^{B,L}$(b), and $\chi_{10}^{B,L}$(c) at $\Delta T=1.0$ MeV with different values of $w$ and $\rho$. }
\end{figure*}

In the left column of Fig.~2, one can found that the general patterns in density plots of $\chi_6^{B,L}$ do not change with varying values of $w$ and $\rho$. So do that of $\chi_8^{B,L}$ and $\chi_{10}^{B,L}$ in the middle and right columns. As the value of $\mu_B$ increases, the density plot of $\chi_{6}^{B,L}$, $\chi_{8}^{B,L}$, and $\chi_{10}^{B,L}$ exhibits alternating negative and positive lobes. The higher the order of susceptibility, the more frequent this alternation becomes, leading to a greater number of sign changes in the susceptibilities.

Comparing each row of Fig.~2, it is clear that the main pattern around the critical point is wider in the $T$ direction in the top row than the other rows, where $w=0.4$ is smaller in the top row than $w=0.8$ in the other two rows. On the other hand, in the $\mu_B$ direction, the main pattern around the critical point is narrower in the bottom row with $\rho=0.4$ than the upper rows where $\rho=0.8$. These effects of $w$ and $\rho$ are consistent with the results with Ref.~\cite{stephanov-prc103}. Smaller $w$ leads to wider critical region in $T$ direction, while smaller $\rho$ leads to narrower critical region in $\mu_B$ direction.

\begin{figure*}[hbt]
	\centering
	\includegraphics[width=0.32\textwidth]{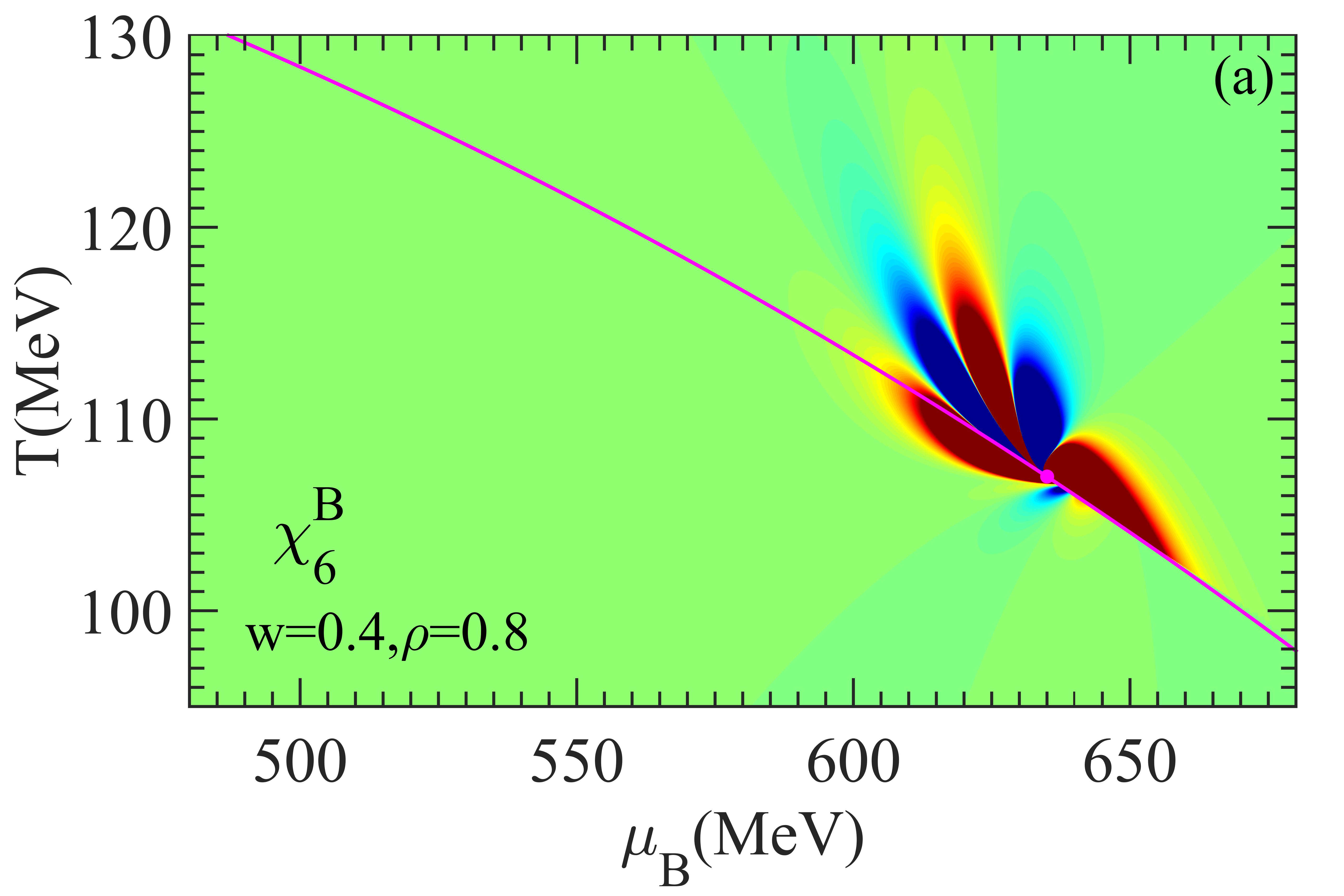}
	\includegraphics[width=0.32\textwidth]{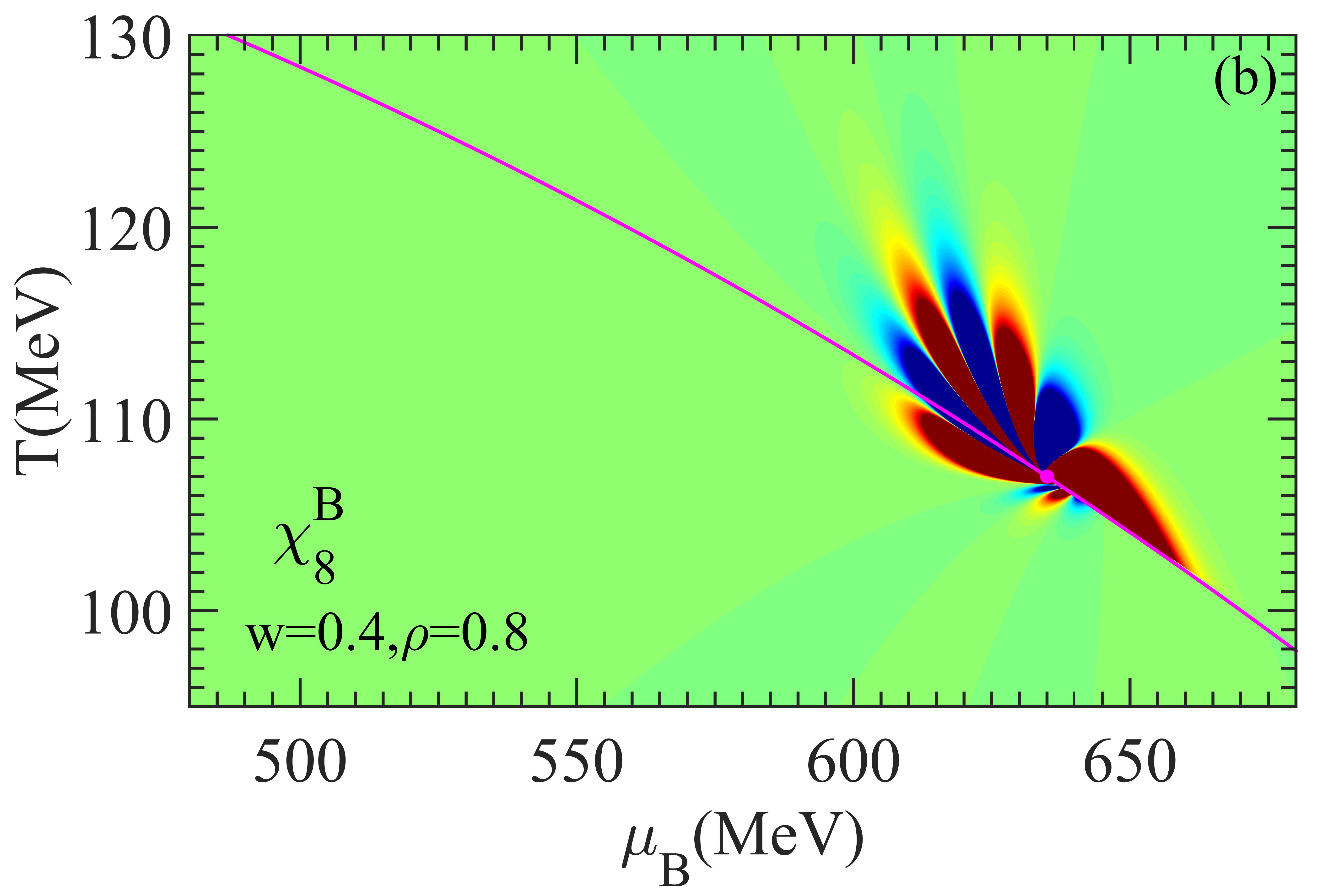}
	\includegraphics[width=0.32\textwidth]{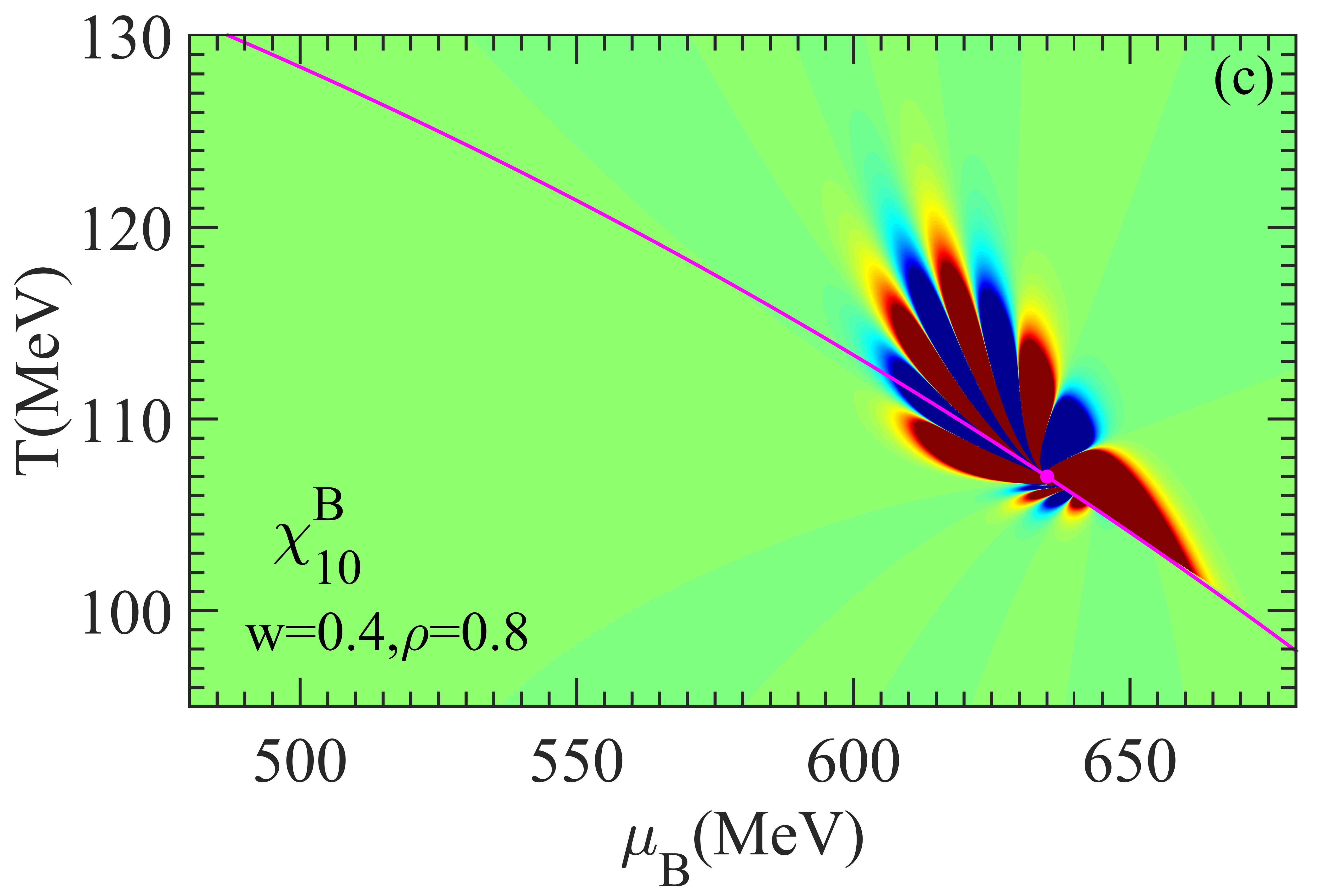}
	\includegraphics[width=0.32\textwidth]{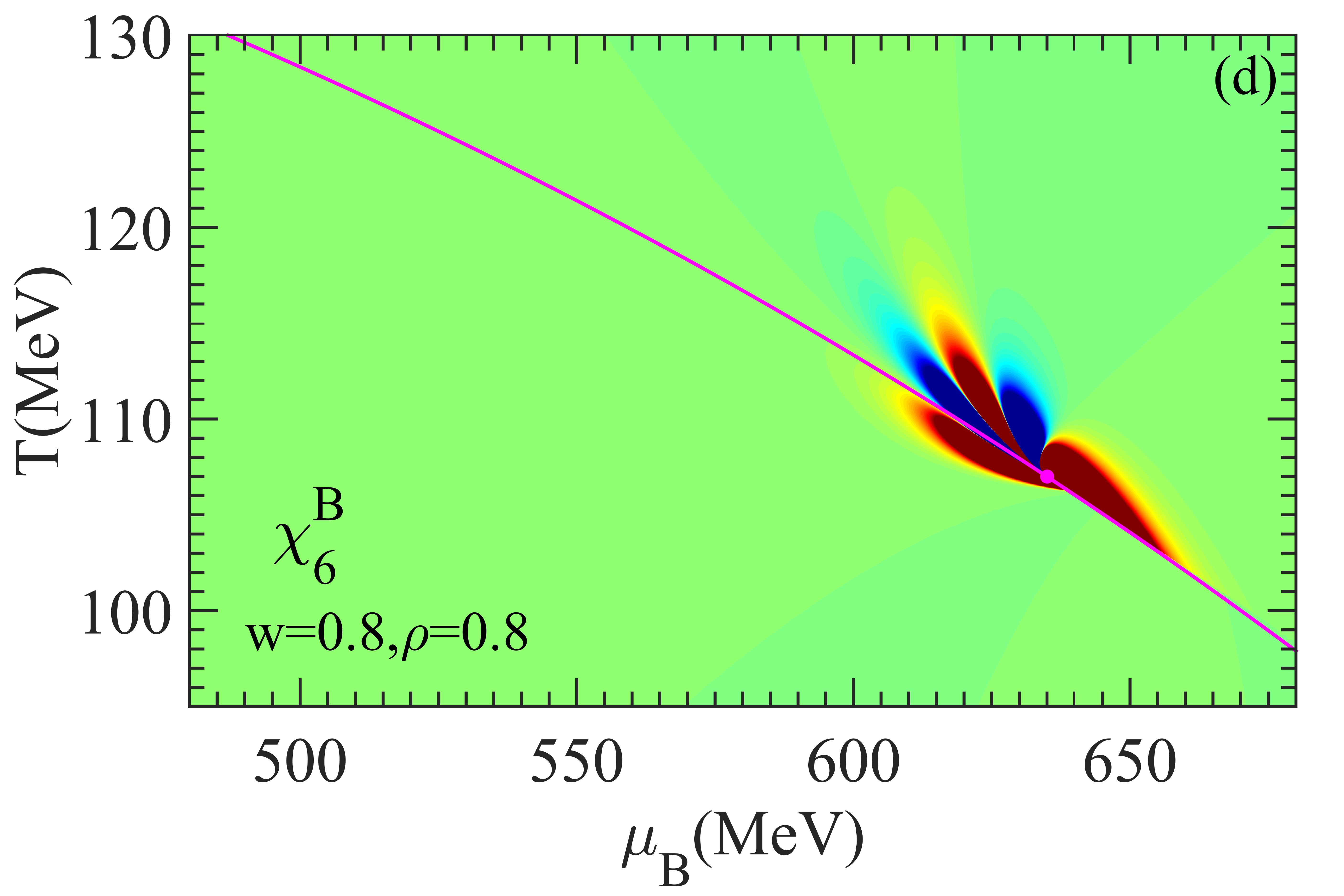}
	\includegraphics[width=0.32\textwidth]{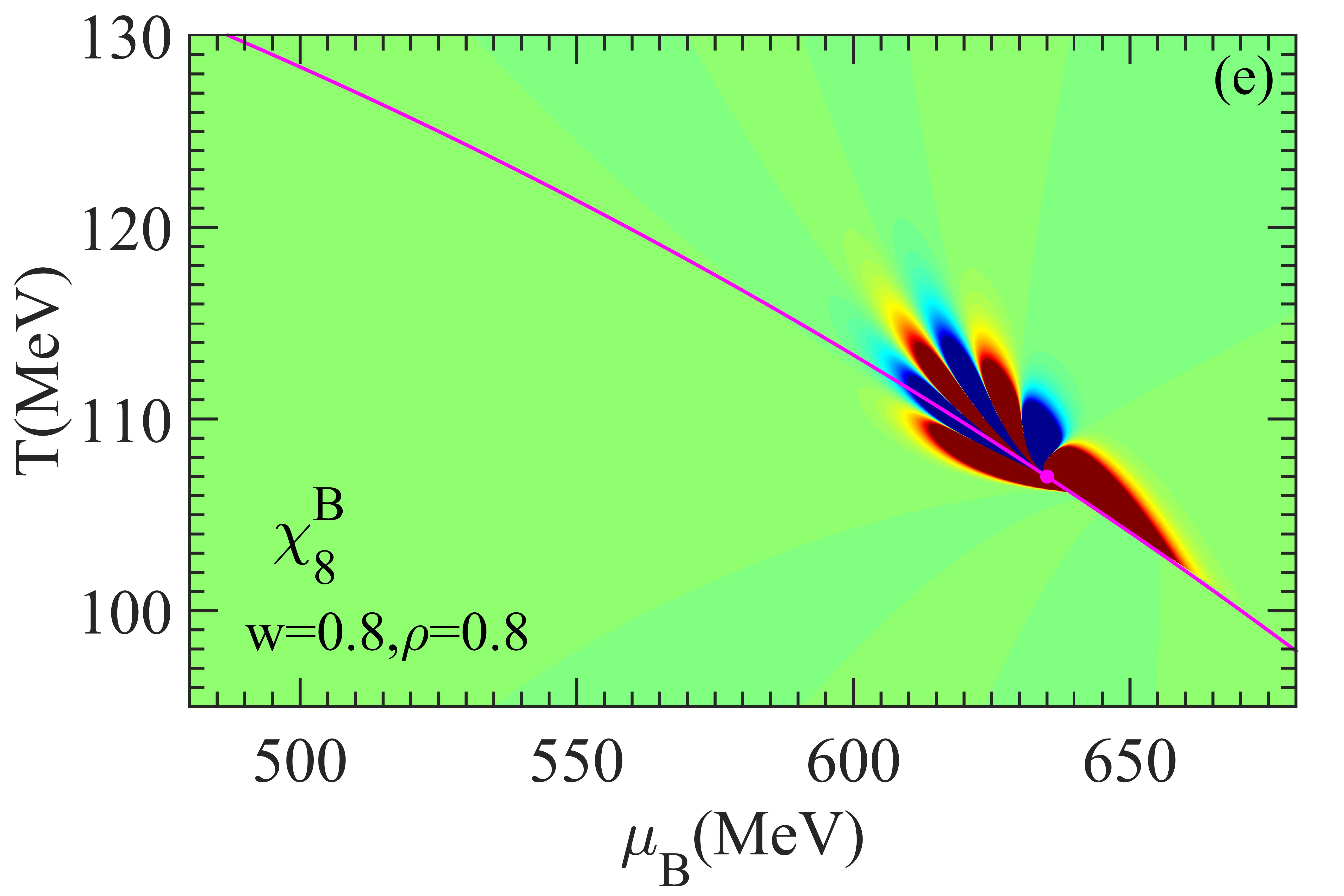}
	\includegraphics[width=0.32\textwidth]{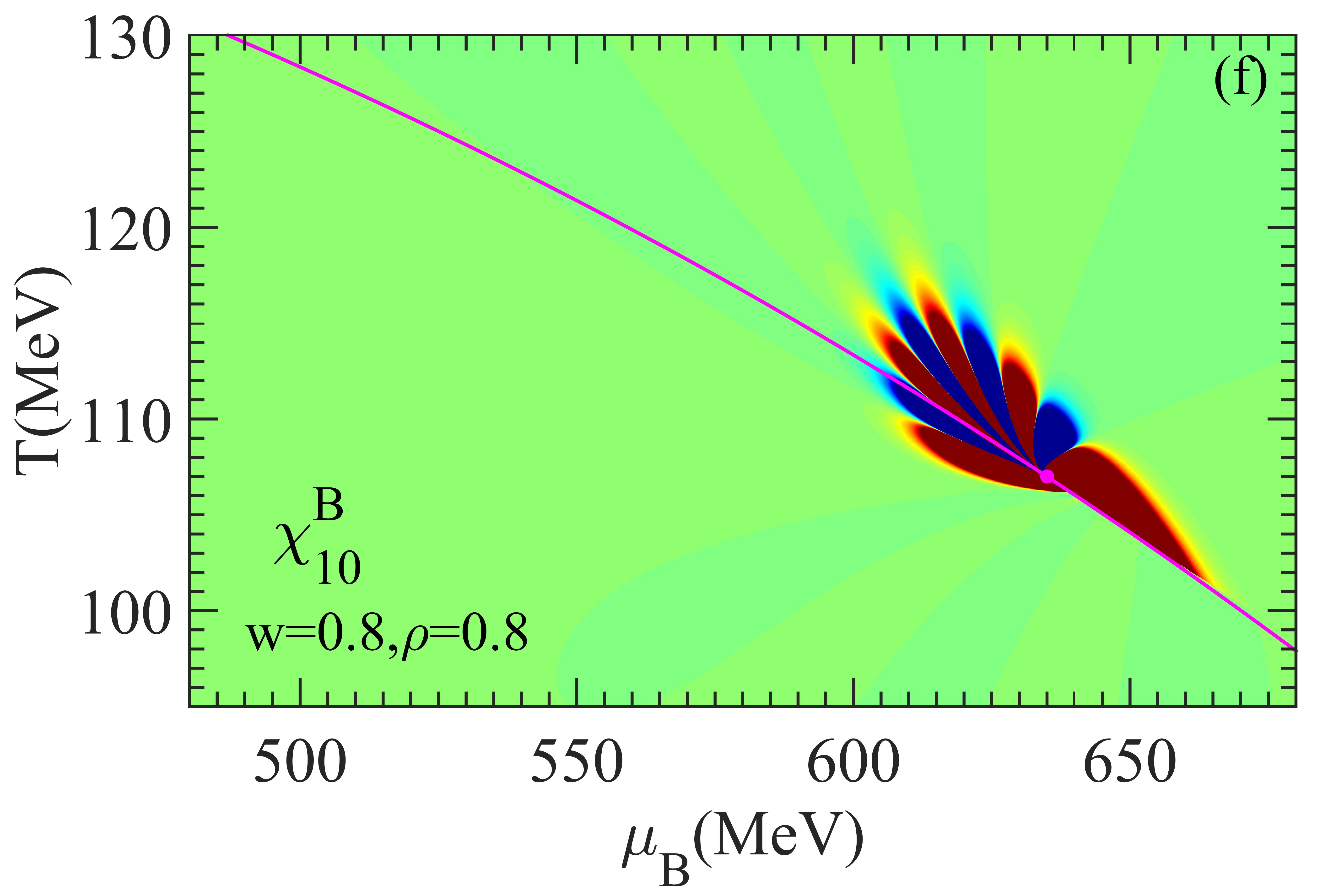}
	\includegraphics[width=0.32\textwidth]{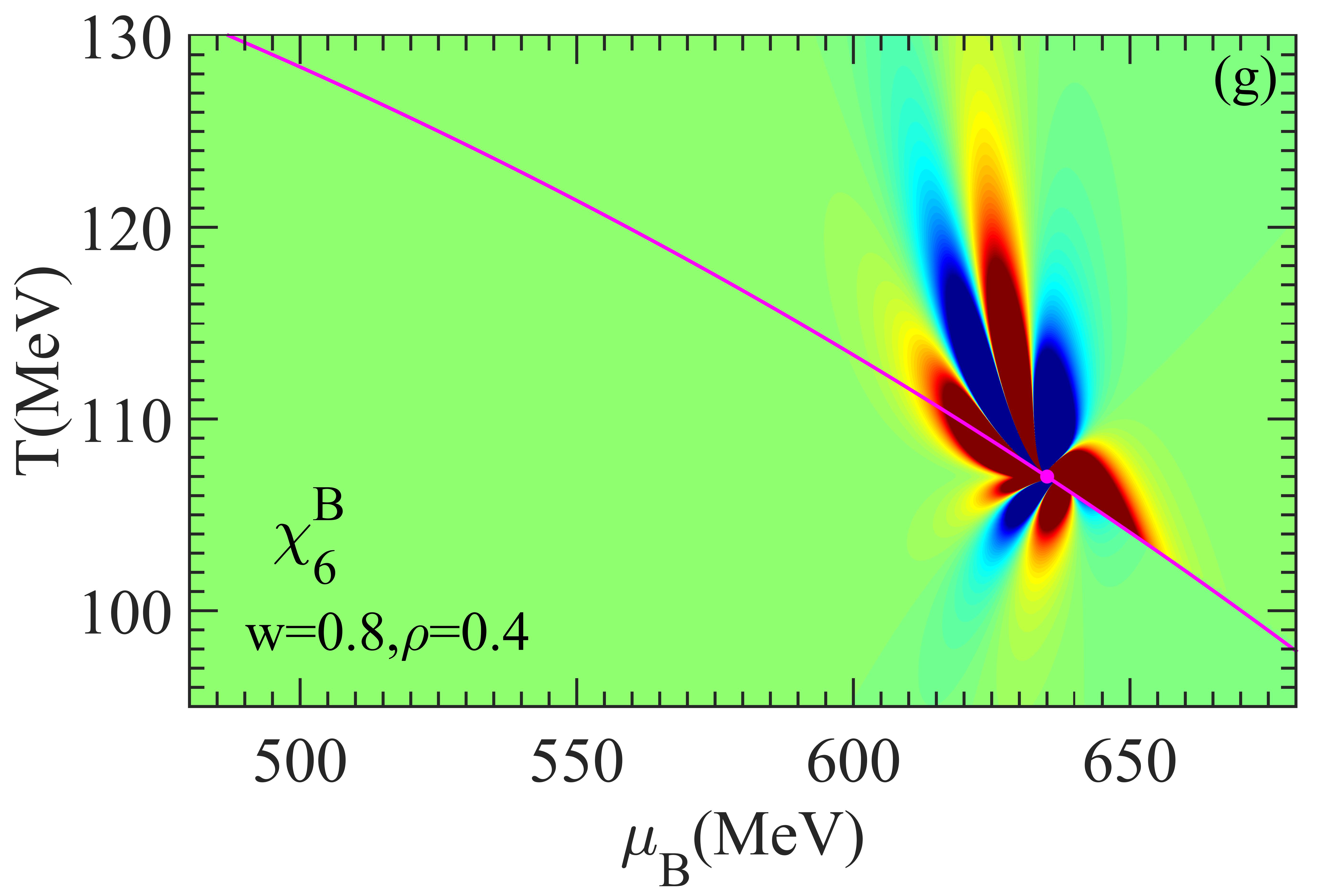}
	\includegraphics[width=0.32\textwidth]{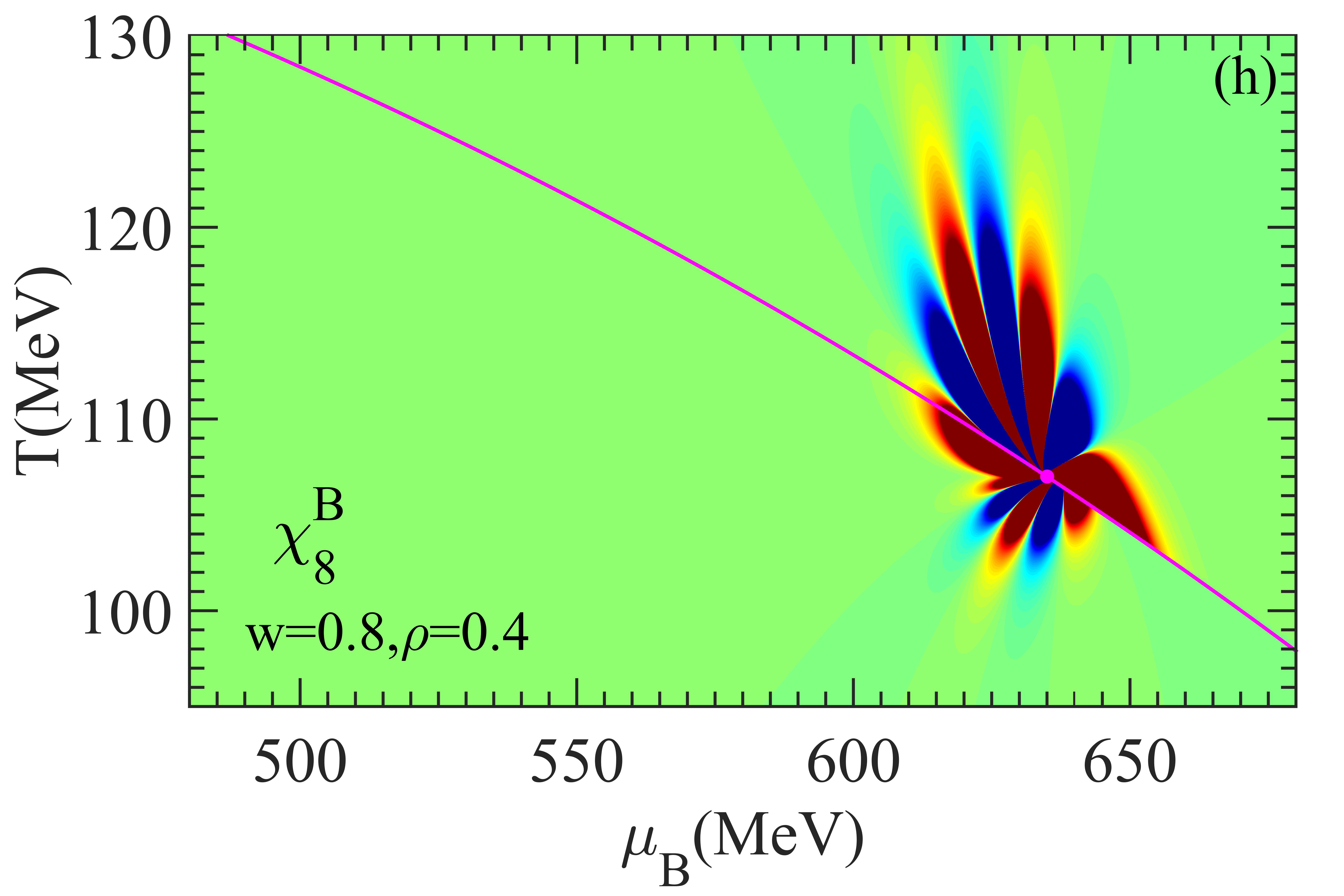}
	\includegraphics[width=0.32\textwidth]{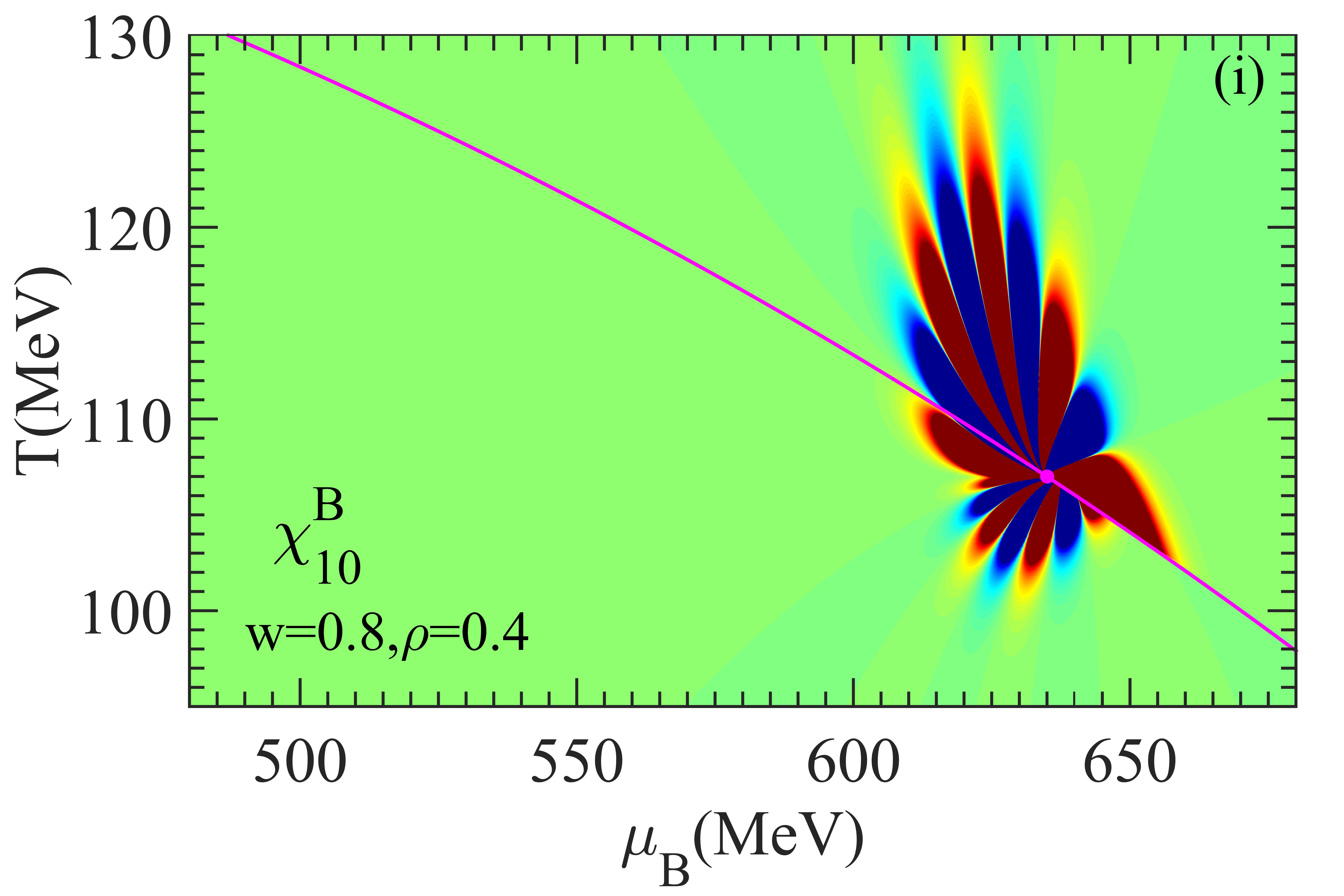}
	
	\caption{\label{Fig. 4}(Color online). Density plots of critical contribution to $\chi_{6}^{B}$, $\chi_{8}^{B}$, and $\chi_{10}^{B}$ in the QCD $T-\mu_B$ phase plane with $w=0.4,\rho=0.8$ (top row), $w=0.8$ and $\rho=0.8$ (middle row) and $w=0.8,\rho=0.4$ (bottom row). The critical point is indicated by a purple dot, while the chiral phase transition line is represented by the solid purple line. The green, yellow and red areas correspond to positive values (the regions where it is the largest and smallest are indicated in red and green, respectively) of the susceptibilities, while the blue ones correspond to negative values (the darker the blue, the larger in magnitude of the susceptibilities). }
\end{figure*}

In order to investigate the $\mu_B$ dependence of net-baryon number susceptibilities and its significance in measuring the energy dependence of net-proton in experiments, we assume a freeze-out curve that approximately parallels the QCD phase transition line as described by Eq.~\eqref{QCD transition line}, but is shifted towards lower temperatures by $\Delta T$, that is
\begin{equation}\label{freeze-out curve}
	T_f(\mu_B)=T_0[1-\kappa(\frac{\mu_B}{T_0})^2-\lambda (\frac{\mu_B}{T_0})^4]-\Delta T.
\end{equation}

The $\mu_B$ dependence of $\chi_{6}^{B,L}$, $\chi_{8}^{B,L}$, and $\chi_{10}^{B,L}$ along the freeze-out curve described by Eq.~\eqref{freeze-out curve} with $\Delta T = 1.0$ MeV are shown in Fig.~3(a), 3(b) and 3(c), respectively. The red, purple, black, cyan and blue curve is for five different combinations of values of $w=0.4,0.8,1.6$ and $\rho=0.4,0.8,1.6$, respectively. The green horizontal and vertical dashed lines show the zero values of the susceptibilities and the net-baryon chemical potential $\mu_{BC}=635$ MeV at the QCD critical point, respectively.

It is clear that the $\mu_B$ dependence of $\chi_{6}^{B,L}$ in Fig.~3(a) has a negative dip followed by a positive peak, when the critical point is approached from the crossover side. So do $\chi_{8}^{B,L}$ in Fig.~3(b) and $\chi_{10}^{B,L}$ in Fig.~3(c). Different values of $w$ and $\rho$ do not change the generic structure of the $\mu_B$ dependence of $\chi_{6}^{B,L}$, and also that of $\chi_{8}^{B,L}$ and $\chi_{10}^{B,L}$.

In each sub-figure, the peak in the red curve is the highest, the peak in the blue curve demonstrates the lowest, while the peaks in the purple, black and cyan curves are of approximately equal height. This observation indicates that as the value of $w$ decreases, there is an increase of peak hight in $\mu_B$ dependence of $\chi_{6}^{B,L}$, $\chi_{8}^{B,L}$ and $\chi_{10}^{B,L}$ along the freeze-out curve, while the influence of $\rho$ on the peak height is small. However, the peak width expands with increasing values of $\rho$. These findings align with the anticipated effects of $w$ and $\rho$ on the critical region.

As the order of the susceptibilities increases, the height of the peak and the depth of the dip both intensify, while their respective widths diminish. Furthermore, the negative dip becomes more pronounced, with the ratio of the depth of the negative dip to the height of the peak growing larger. For instance, this ratio is approximately $0.163$, $0.275$, and $0.361$ for the red curve presented in Fig.~3(a), 3(b), and 3(c), respectively, indicating a trend towards a more significant contrast between the peak and the dip with higher-order susceptibilities.

Although the density plots of $\chi_6^{B,L}$, $\chi_8^{B,L}$ and $\chi_{10}^{B,L}$ in Fig.~2 suggest that the $\mu_B$ dependence of these susceptibilities should undergo multiple sign changes as their order increases, only the positive lobe immediately below the critical point and its nearest negative lobe lead to the prominent peak and dip in Fig.~3. The magnitudes of the values of other lobes are notably smaller compared to that of the two lobes closest to the critical point, making them difficult to be observed in the $\mu_B$ dependence.

\begin{figure*}[hbt]
	\centering
	\includegraphics[width=0.32\textwidth]{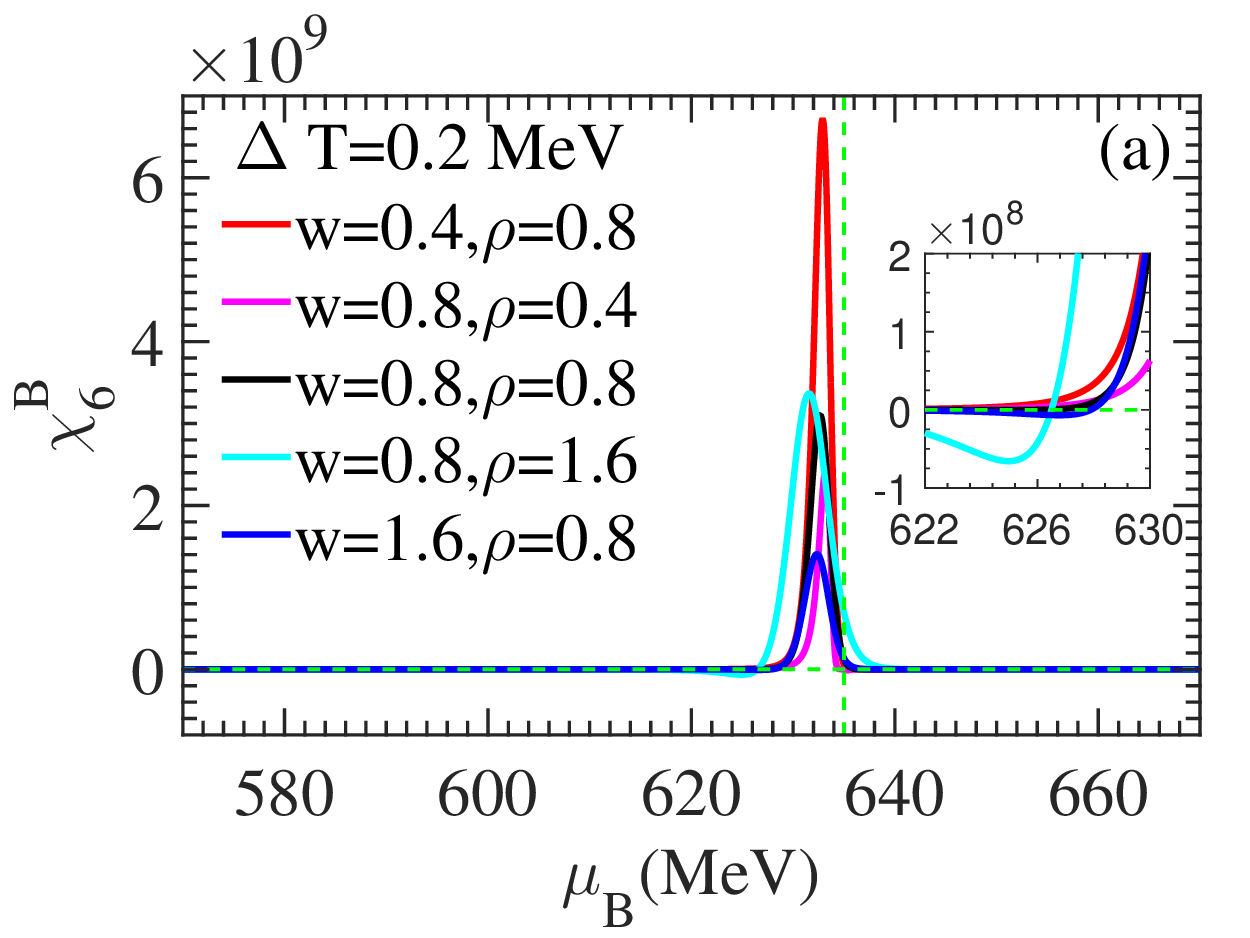}
	\includegraphics[width=0.32\textwidth]{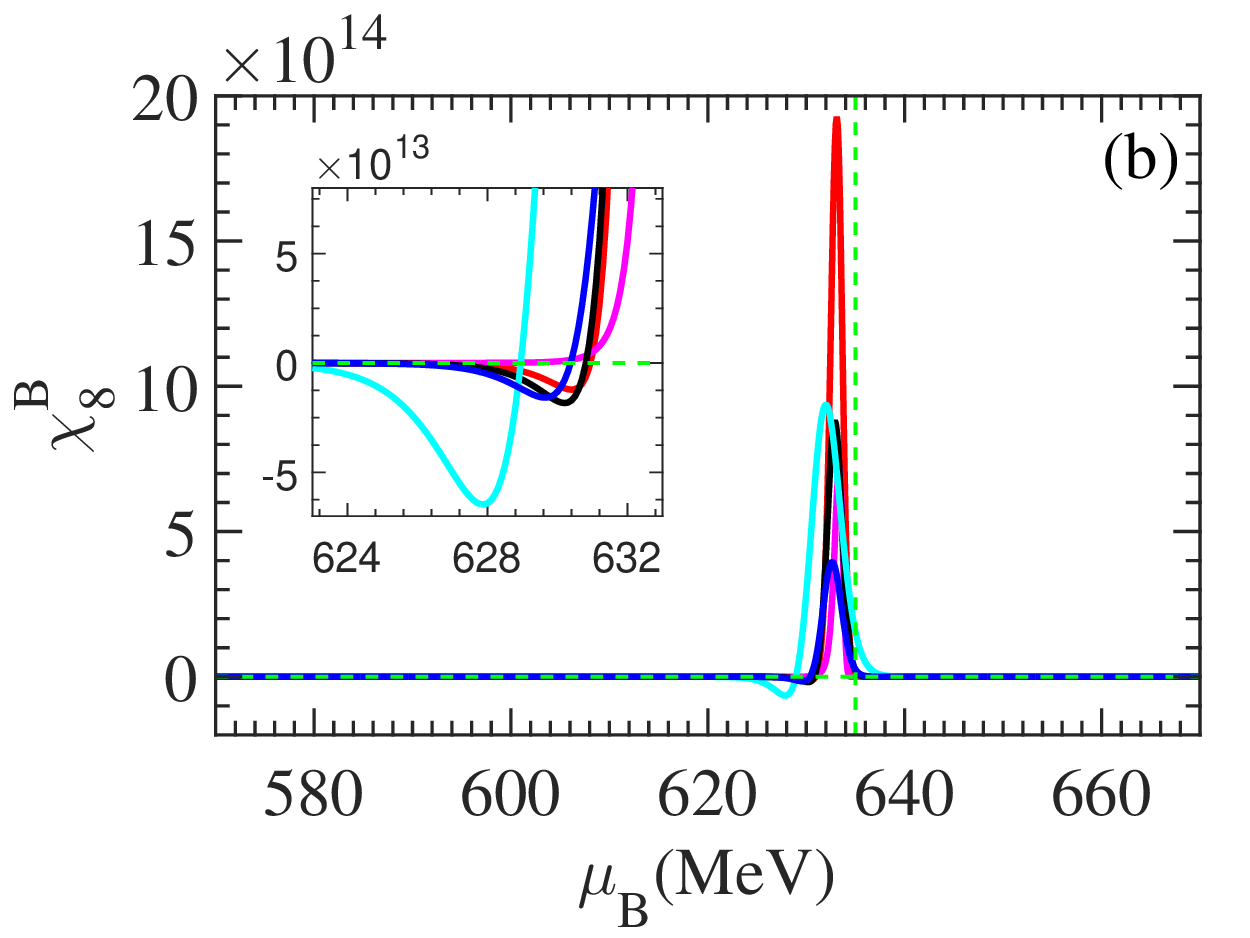}
	\includegraphics[width=0.32\textwidth]{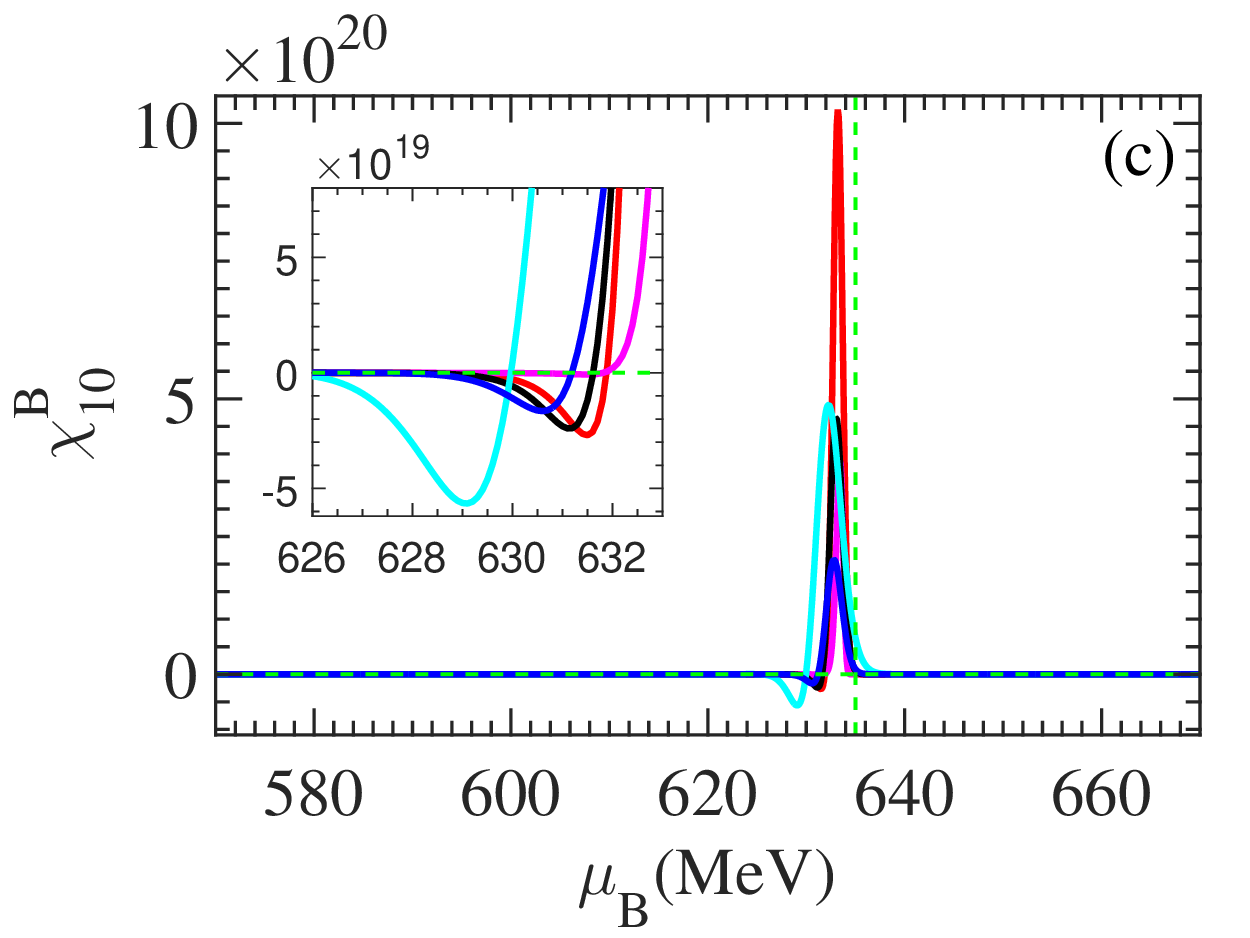}
	\includegraphics[width=0.32\textwidth]{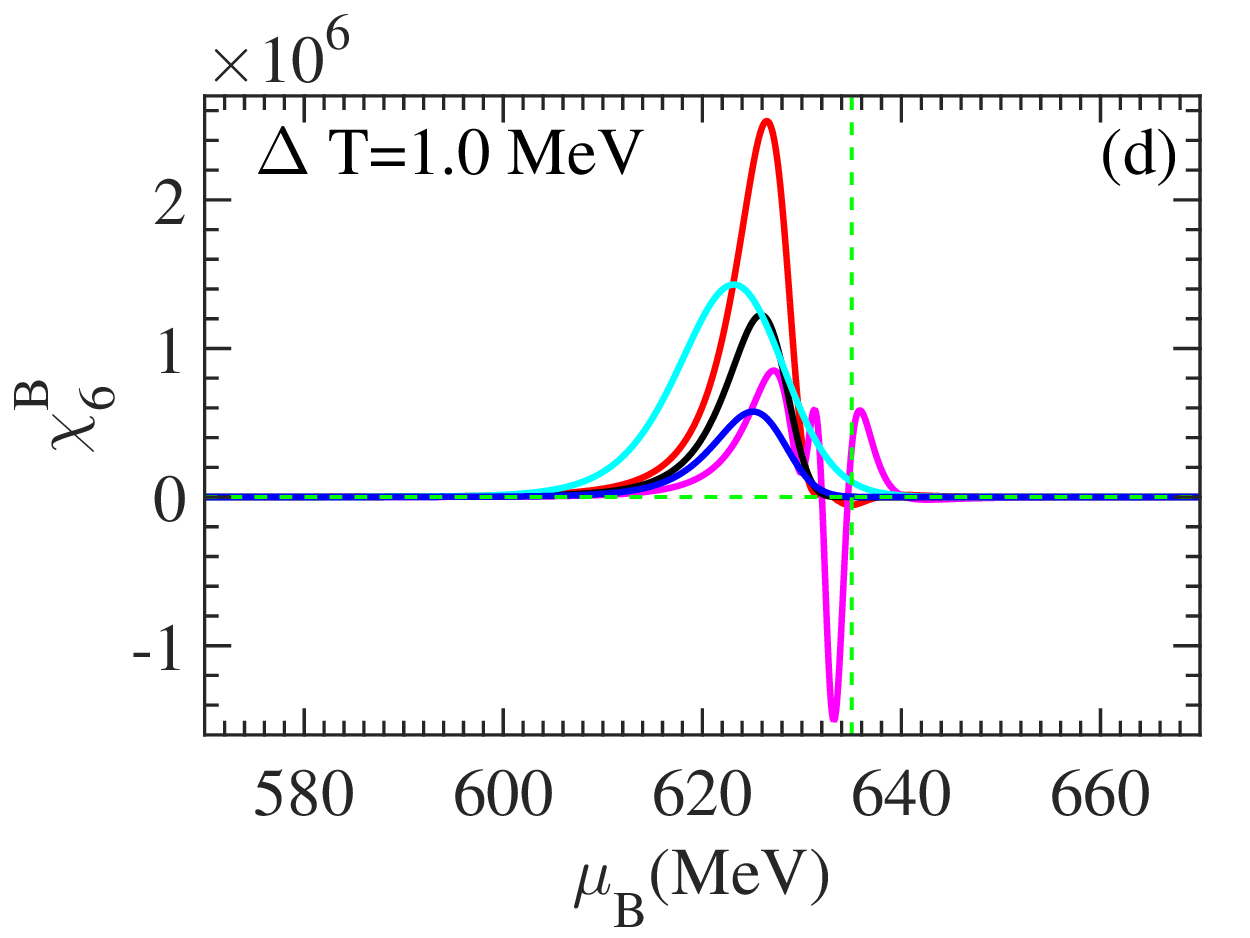}
	\includegraphics[width=0.32\textwidth]{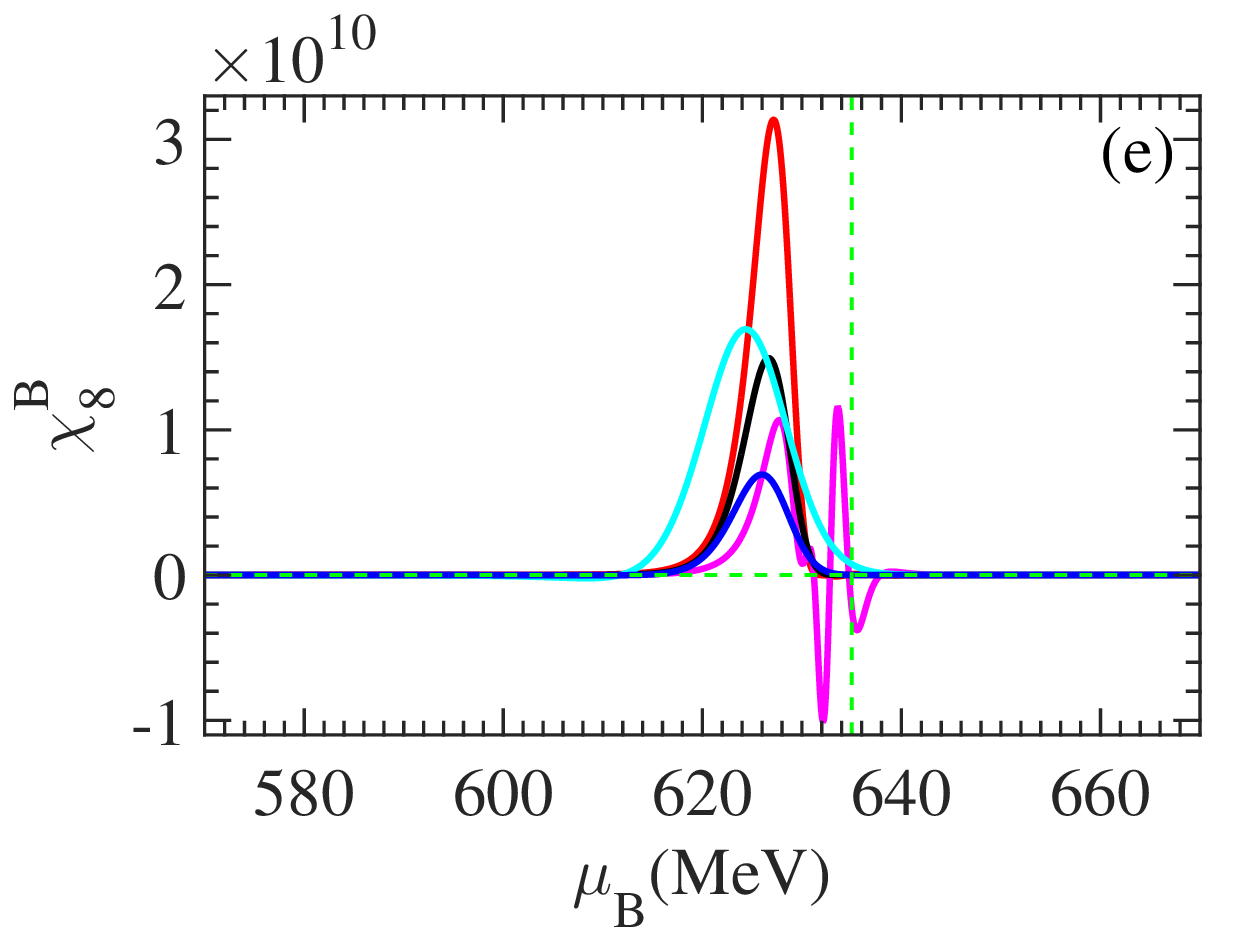}
	\includegraphics[width=0.32\textwidth]{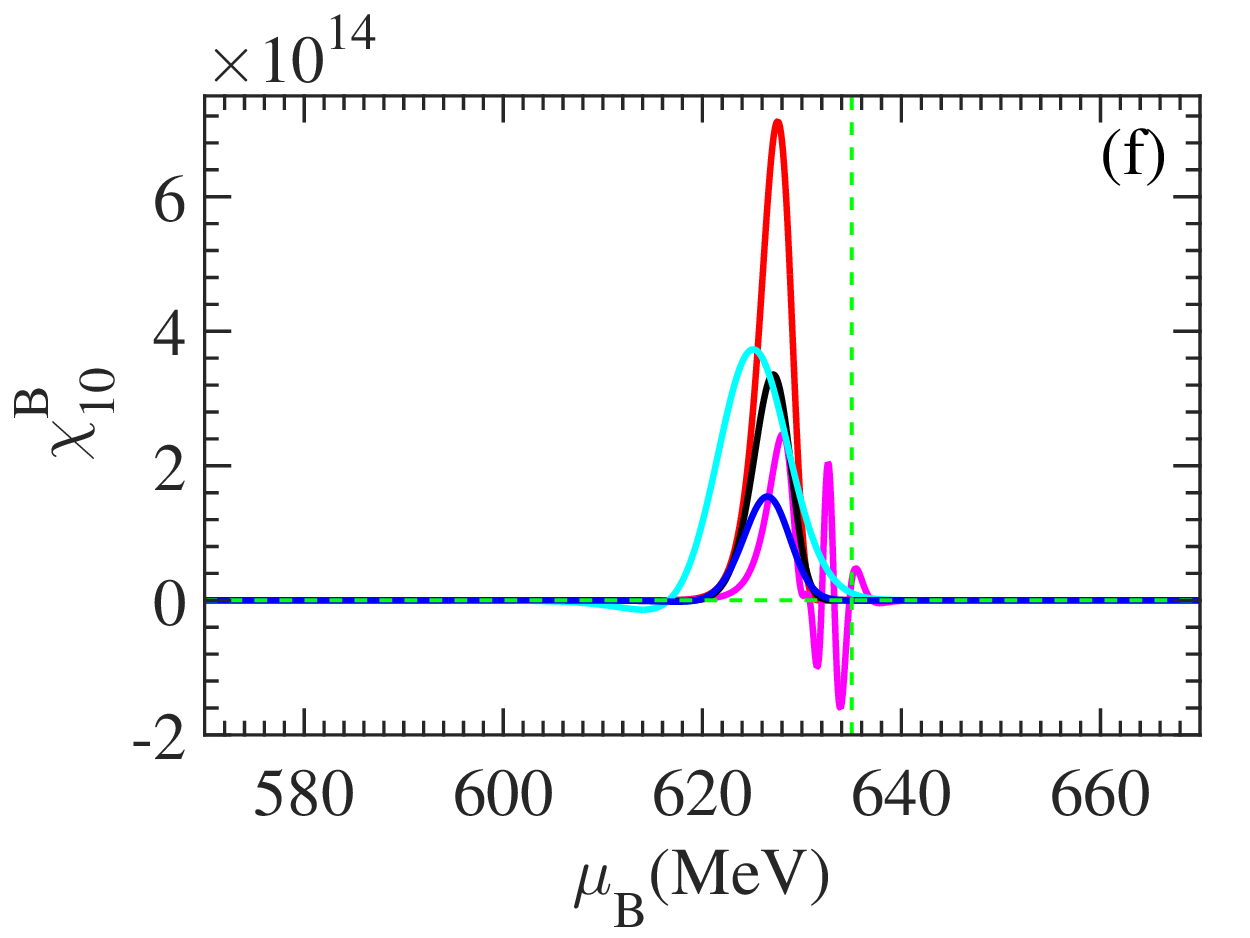}
	\includegraphics[width=0.32\textwidth]{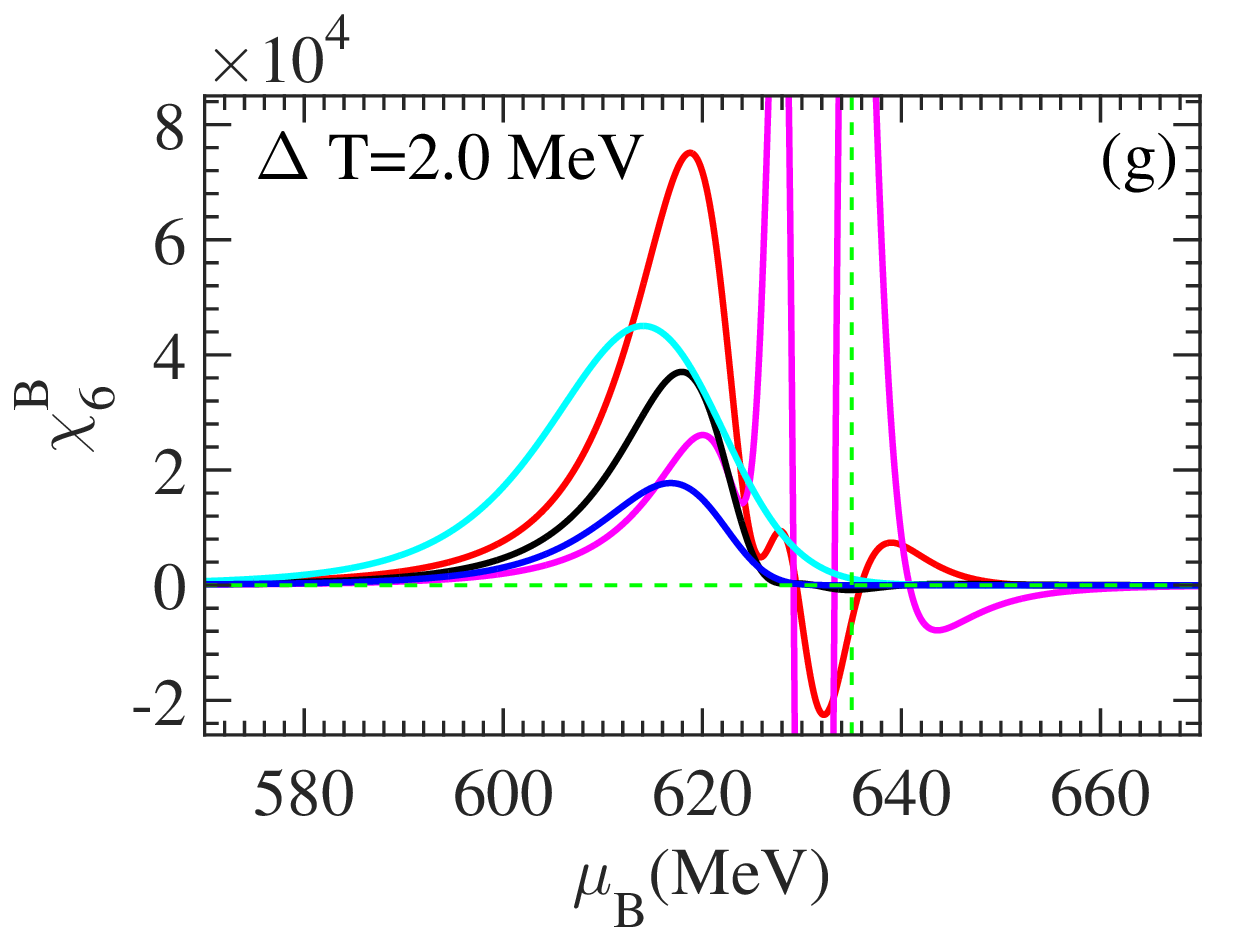}
	\includegraphics[width=0.32\textwidth]{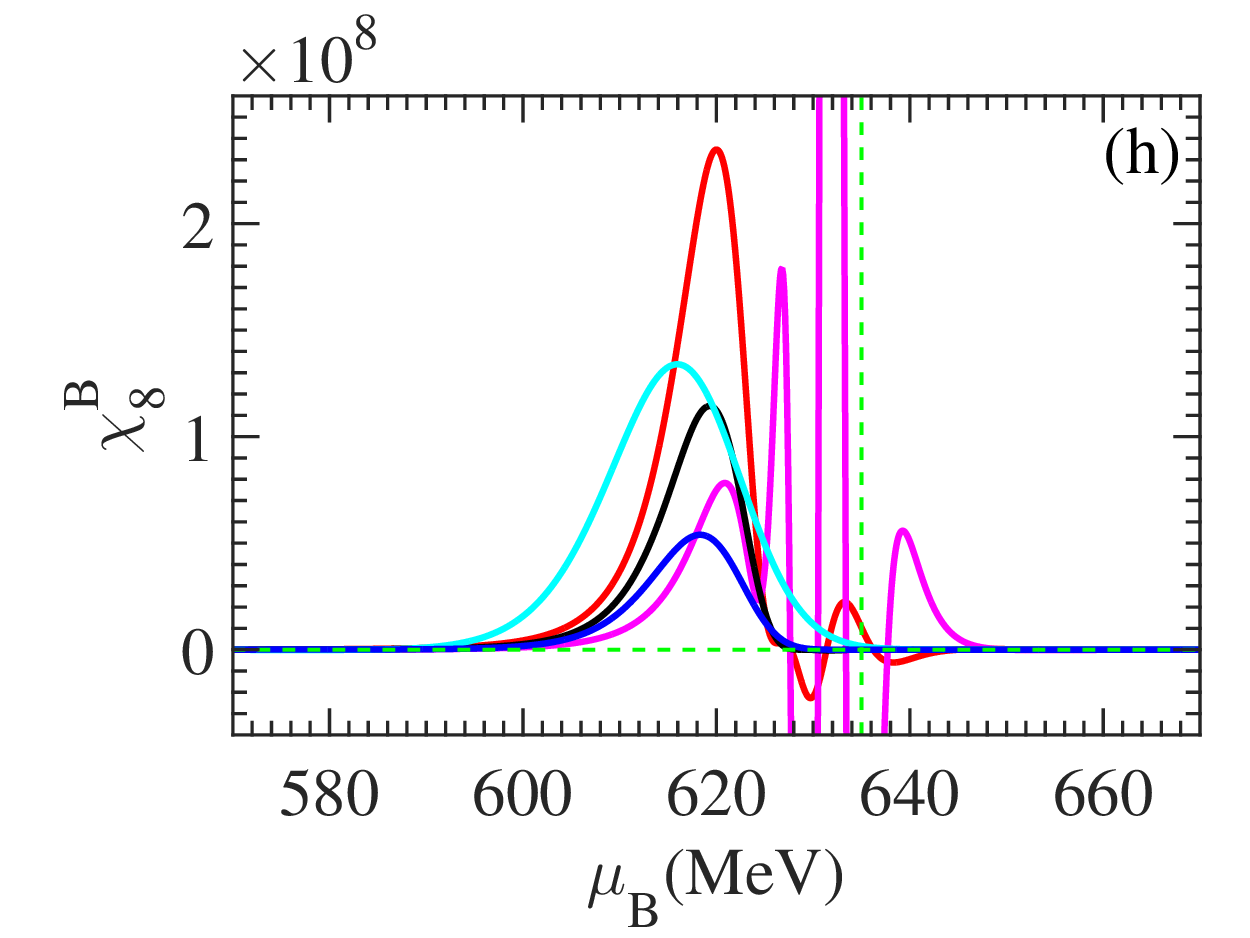}
	\includegraphics[width=0.32\textwidth]{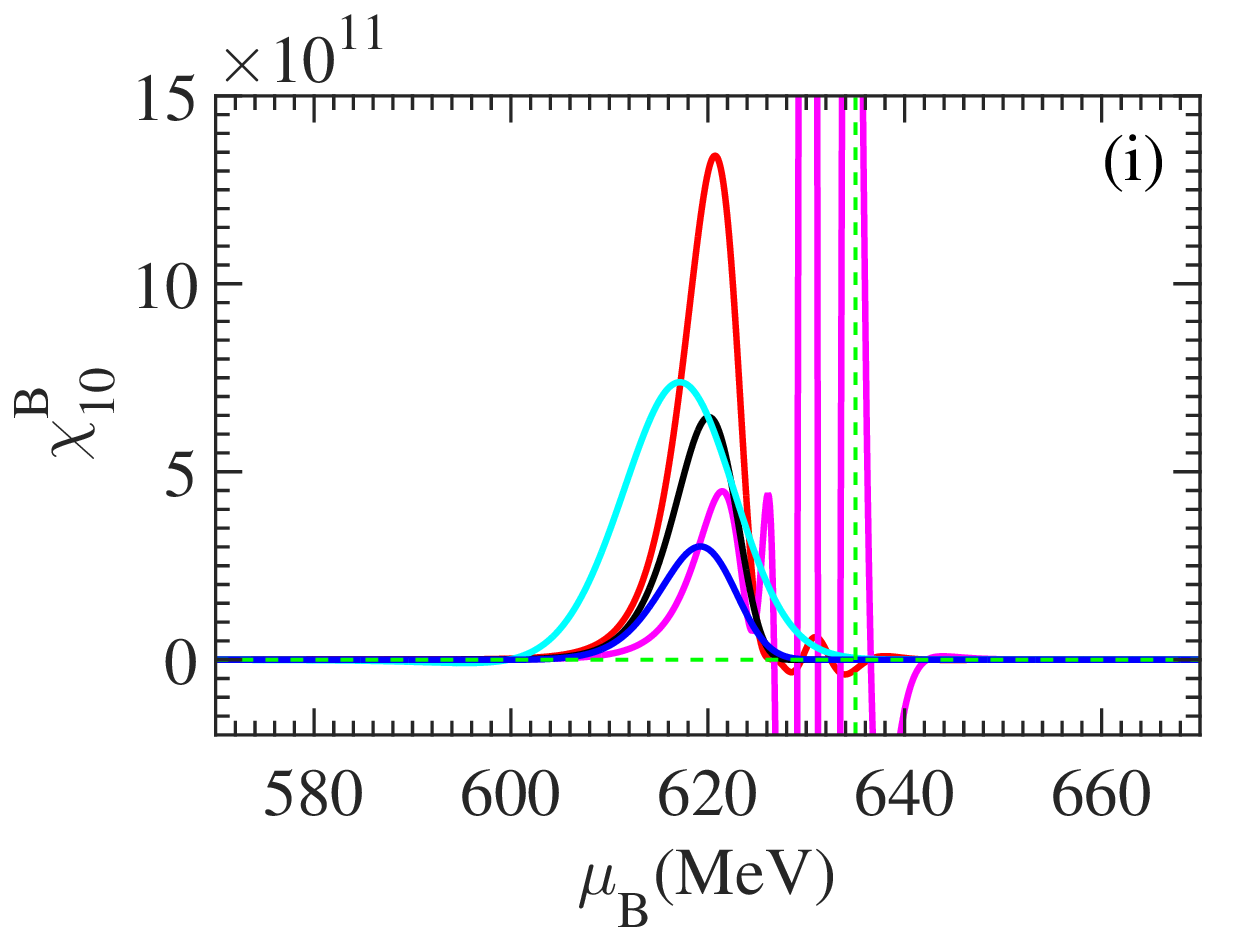}
	\caption{\label{Fig. 5}(Color online). $\mu_B$ dependence of $\chi_{6}^{B}$, $\chi_{8}^{B}$, and $\chi_{10}^{B}$ at $\Delta T=0.2$ MeV (top row), $\Delta T=1.0$ MeV (middle row) and $\Delta T= 2.0$ MeV (bottom row) with different values of $w$ and $\rho$.}
\end{figure*}

\section{Sub-leading critical contribution to the behavior of susceptibilities of net-baryon number }

In comparison to the sub-leading critical contributions to the net-baryon number susceptibilities, i.e., the terms those $k \neq 0$ in Eq.~\eqref{susceptibilities}, the leading singular contribution will be suppressed by power of $\sin \alpha_1/\sin \alpha_2$ in the case $\alpha_1$ is small, while $\alpha_2$ is not small~\cite{stephanov-prc103}. It is exactly the case for the common choice of the mapping from Ising variables to the QCD $T-\mu_B$ phase plane, such as $\alpha_1=10.8^\circ$ and $\alpha_2=100.8^\circ$ in this paper.

The density plots of the critical contribution to $\chi_6^B$, $\chi_8^B$ and $\chi_{10}^B$ are shown in Fig.~4, utilizing the same parameters as those employed in Fig.~2. I.e., $w=0.4$, $\rho=0.8$ for the top row, $w=0.8$, $\rho=0.8$ for the middle row, and $w=0.8$, $\rho=0.4$ for the bottom row.

In Fig.~4, the color schemes for $\chi_{2n}^{B}$ maintain consistency with the corresponding order of $\chi_{2n}^{B,L}$ presented in Fig.~2. The purple curve shows the QCD phase transition line, as defined by Eq.\eqref{QCD transition line}, while the purple dot marks the critical point.

Comparing the corresponding sub-figure in Fig.~4 with that in Fig.~2, it is clear that the sub-leading critical contribution alters the pattern away from the critical point. We summarize three points as follows.

Firstly, the area occupied by the primary pattern (consist of the red and dark blue lobes) around the critical point in each sub-figure of Fig.~4 is larger than that in the corresponding sub-figure of Fig.~2.

Secondly, more lobes located in the area above the phase transition line, while less lobes occur under the phase transition line as showed in the middle row of Fig.~4, where the values of $w$ and $\rho$ are equal to $0.8$. Additional lobes appear under the critical point at larger $\mu_B$ side, but their magnitudes are very small (the colors of the additional lobes are green or light blue). When it is extremely close to the phase transition line, negative values of the susceptibilities (dark blue lobes) can be observed under the phase transition line at lower $\mu_B$ side. Although not explicitly presented in this paper, it should be pointed out that for values of $w$ and $\rho$ that exceed $0.8$, the density plots of $\chi_6^B$, $\chi_8^B$ and $\chi_{10}^B$ are similar to that of the corresponding susceptibilities shown in the middle row of Fig.~4, respectively. The numbers of red and dark blue lobes in the density plots keep the same. The main effect of increasing values of $w$ and $\rho$ is the compression of the primary pattern in the $T$ direction and the stretch of that in the $\mu_B$ direction.

Thirdly, for smaller values of $w$ or $\rho$, as showed in the top and bottom rows of Fig.~4 respectively, the additional lobes emerging below the critical point become obvious. Their colors turn to red and dark blue, which may result in additional positive peak and negative dips in the $\mu_B$ dependence of susceptibilities along the freeze-out curve. The higher the order of the susceptibility, the more additional lobes it has and more times its sign changes.

To study the $\mu_B$ dependence of $\chi_6^B$, $\chi_8^B$ and $\chi_{10}^B$ in detail, except $\Delta T=1.0$ MeV, we choose another two freeze-out curves described in Eq.~\eqref{freeze-out curve}, where $\Delta T$ is equal to $0.2$ MeV and $2.0$ MeV, respectively. $\mu_B$ dependence of the susceptibilities along the three different freeze-out curves are shown in the top, middle and bottom rows of Fig.~5, respectively. In each sub-figure, the red, purple, black, cyan and blue curve is for five different combinations of values of $w=0.4,0.8,1.6$ and $\rho=0.4,0.8,1.6$, respectively. The green horizontal and vertical dashed lines show the zero values of the susceptibilities and the net-baryon chemical potential $\mu_{BC}=635$ MeV at the QCD critical point, respectively.

In the top row of Fig.~5, $\Delta T=0.2$ MeV, the freeze-out curve is very close to the phase transition line, the negative dip in the $\mu_B$ dependence of the susceptibilities is likely to appear based on their density plots shown in Fig.~4. To clearly observe the negative dip, partial enlarged details are provided for Fig.~5(a), Fig.~5(b) and 5(c). It is clear that a negative dip followed by a positive peak can be observed in the cyan curve; similarly, in Fig.~5(b) and 5(c), this structure can also be observed in the black, blue, and red curves as one approaches the critical point from the crossover side. However, compared to Fig.~3, the negative dip is less pronounced; for instance, for the cyan curve in Fig.~5(a), 5(b), and 5(c), respectively, ratios of depth of negative dips to height of peaks are $0.016$, $0.058$, and $0.099$, smaller than that in the red curves of Fig.~3.

For the red and purple curves in each sub-figure in the top row of Fig.~5, additional negative dips or positive peaks have not been observed at larger $\mu_B$ side. This is because the magnitude of values of the additional lobes in the top and bottom rows of Fig.~4 is too small to form a pronounced dip or peak. The original red lobe dominates the main peak structure when the freeze-out curve is very close to the phase transition line.

From the top row to the bottom row in Fig.~5, the freeze-out curve is located more and more distant from the phase transition line. The negative dip at lower $\mu_B$ side fades away in the red, black, cyan, and blue curves. At the same time, the additional positive peaks and negative dips in the red and purple curve at larger $\mu_B$ side emerge and become more pronounced.

The $\mu_B$ dependence of $\chi_6^B$, $\chi_8^B$ and $\chi_{10}^B$ slightly differs from that of the fourth-order susceptibilities of net-baryon number when $\alpha_1$ is small and $\alpha_2$ is not small. In this scenario, the negative dip disappears in the fourth-order susceptibility for all values of $w$ and $\rho$ as reported in Ref.~\cite{stephanov-prc103}. While for $\chi_6^B$, $\chi_8^B$ and $\chi_{10}^B$ , the existence of the negative dip is dependent on the values of $w$ and $\rho$, as well as the distance between the freeze-out curve with the phase transition line. When both $w$ and $\rho$ are large, negative dip followed by positive peak can be observed when the critical point is approached from the crossover side, provided that the freeze-out curve is very close to the phase transition line; however, the magnitude of this negative dip remains minimal. For small values of $w$ and $\rho$, additional negative dips will emerge and become pronounced as the freeze-out curve is further away from the phase transition line.

While the presence of negative dips in the $\mu_B$ dependence of $\chi_6^B$, $\chi_8^B$ and $\chi_{10}^B$ relies on mapping parameters and the distance between the freeze-out curve and the phase transition line, a positive peak consistently emerges in the $\mu_B$ dependence of susceptibilities as the critical point is approached from the crossover side. This occurrence remains unaffected by values of $w$ and $\rho$, as well as by the distance between the freeze-out curve and phase transition line. Specifically, this positive peak is observed at the left side of vertical green dashed line in every sub-figure within Fig.~5. In contrast to negative dips, this positive peak structure in net-baryon number susceptibilities represents a more robust characteristic of the critical point.

\section{Summary}

Assuming the equilibrium of the QCD system, the results from the three-dimensional Ising model can be mapped to that of QCD based on the universality of critical behavior. Applying the common choice for the mapping, i.e. the Ising magnetic field is orthogonal to the temperature, we have investigated the critical behavior of sixth-, eighth- and tenth-order susceptibilities of the net-baryon number. Both the leading critical contribution as well as sub-leading critical contribution from the Ising model are discussed.

When just taking the leading critical contribution into account, the general patterns of density plots for susceptibilities of the same order remain consistent across varying values of mapping parameters $w$ and $\rho$. The higher the order of the susceptibility, the more lobes in the pattern of the density plot and more times of the sign changes. The common feature is the occurrence of a negative dip followed by a positive peak in the $\mu_B$ dependence of the sixth-, eighth-, and tenth-order susceptibilities of net-baryon number when the critical point is approached from the crossover side.

The sub-leading critical contribution significantly affect the behavior of the susceptibilities. In $\mu_B$ dependence of the susceptibilities, the emergence of negative dips is not an absolute phenomenon; rather, it is dependent on the values of the mapping parameters, as well as the distance between the freeze-out curve and the phase transition line.

In comparison to negative dip in the $\mu_B$ dependence of generalized susceptibilities of net-baryon number, the positive peak structure is a more robust feature of the critical point, which is unaffected by values of $w$ and $\rho$, as well as by the distance between the freeze-out curve and phase transition line, or sub-leading critical contribution.

\vskip 0.5cm
This work is supported by Research Start-up Fund Project of Chengdu University (X2110) and I acknowledge fruitful discussions with Xiaofeng Luo and Lizhu Chen.

\ed